\documentclass[12pt]{article}

\usepackage[dvips]{graphicx}
\usepackage{epsf}
\usepackage{cite}
\usepackage{amssymb}

\makeatletter
\@addtoreset{equation}{section}
\makeatother

\begin{document}


\begin{titlepage}
 
$\mbox{ }$
\begin{flushright}
 \begin{tabular}{l}{\normalsize {\large {\large }}}
  KEK-TH-1284\\
 \end{tabular}
\end{flushright}

\vspace*{0mm}

\begin{Large}
 \vspace{20mm}
 \begin{center}
  {\bf ${\cal N}=4$ Supersymmetric Yang-Mills on $S^3$ \\
  in\\
  Plane Wave Matrix Model 
  at Finite Temperature}\\
 \end{center}
\end{Large}

\vspace{10mm}


\begin{center}
 Yoshihisa K{\sc itazawa}$^{1),2)}$
 \footnote{E-mail address: kitazawa@post.kek.jp}
 and
 Koichiro M{\sc atsumoto}$^{1)}$
 \footnote{E-mail address: kmatsumo@post.kek.jp}\\
 \vspace{5mm}
 $^{1)}$
 {\it Institute of Particle and Nuclear Studies}\\
 {\it High Energy Accelerator Research Organization (KEK)}\\
 {\it Tsukuba, Ibaraki 305-0801, Japan}\\
 $^{2)}$
 {\it Department of Particle and Nuclear Physics}\\
 {\it The Graduate University for Advanced Studies (SOKENDAI)}\\
 {\it Tsukuba, Ibaraki 305-0801, Japan}\\
\end{center}

\vspace{20mm}

 
\begin{abstract}
 We investigate the large $N$ reduced model of gauge theory on a
 curved spacetime through the plane wave matrix model.
 We formally derive the action of the ${\cal N}=4$ supersymmetric Yang-Mills
 theory on ${\mathbb R} \times S^3$ from the plane wave
 matrix model in the large $N$ limit.
 Furthermore, we evaluate the effective action of the plane wave matrix
 model up to the two-loop level at finite temperature.
 We find that the effective action is consistent with the free energy of
 the ${\cal N}=4$ supersymmetric Yang-Mills theory on $S^3$ at high
 temperature limit where the planar contributions dominate.
We conclude that the plane wave matrix model can be used as a large $N$
reduced model to investigate
nonperturbative aspects of the ${\cal N}=4$ supersymmetric Yang-Mills theory
 on ${\mathbb R} \times S^3$.
\end{abstract}

\end{titlepage}


\section{Introduction}

Matrix models are strong candidates for the non-perturbative formulation
of the superstring theory.
For example, the BFSS matrix model is the non-perturbative formulation
of the M-theory which is the strongly coupled limit of the
type-I\hspace{-.1em}IA superstring theory
\cite{banks_fischler_shenker_susskind-1997} and the IKKT matrix model was
proposed as the non-perturbative formulation of the
type-I\hspace{-.1em}IB superstring theory
\cite{ishibashi_kawai_kitazawa_tsuchiya-1997,aoki_iso_kawai_kitazawa_tsuchiya_tada-1999}.
Originally, these models were constructed on flat spacetime backgrounds.
So, we have problems whether these models can describe curved
spacetime, and include symmetries of the general relativity: the
diffeomorphism and the local Lorentz invariance.

In 2002, fuzzy homogeneous spaces are constructed
using the IKKT matrix model \cite{y_kitazawa-2002}.
The homogeneous spaces are constructed as $G/H$ where $G$ is a Lie group
and $H$ is a closed subgroup of $G$. The effective actions of the gauge theory
on homogeneous spaces have been investigated for
a fuzzy $S^2$ \cite{imai_kitazawa_takayama_tomino-2003}, a fuzzy $S^2
\times S^2$
\cite{imai_kitazawa_takayama_tomino-2004,imai_takayama-2004}, a fuzzy
$S^2 \times S^2 \times S^2$ \cite{kaneko_kitazawa_tomino-2005} and a
fuzzy $CP^2$ \cite{kaneko_kitazawa_tomino-2006}.
When a background field is assigned to bosonic matrices in the IKKT matrix
model, the stability of this matrix configuration can be examine by
investigating the behavior of the effective action under the change of
some parameters of the background.
By these investigations, we have found that the IKKT matrix model favors the
configurations of the four-dimensionality.
The same conclusion has been obtained also by various approaches \cite{aoki_iso_kawai_kitazawa_tada-1998,nishimura_vernizzi-2000a,nishimura_vernizzi-2000b,anagnostopoulos_nishimura-2002,nishimura_sugino-2002,kawai_kawamoto_kuroki_matsuo_shinohara-2002,kawai_kawamoto_kuroki_matsuo_shinohara-2003}.

Recently, there were interesting developments on the construction of 
curved spacetime by matrix models.
Hanada, Kawai and Kimura introduced a new interpretation on the IKKT
matrix model in which covariant derivatives on any $d$-dimensional
spacetime can be described in terms of $d$  bosonic matrices
in the IKKT matrix model \cite{hanada_kawai_kimura-2005}.
In this interpretation, the Einstein equation follows from the equation
of the IKKT matrix model, and symmetries of the diffeomorphism and
the local Lorentz transformation are included in the unitary symmetry of
the IKKT matrix model.

On the other hand, the formal equivalences between supersymmetric Yang-Mills
theories on curved spacetime and a matrix model is shown by
Ishiki, Shimasaki, Takayama and Tsuchiya
\cite{ishiki_shimasaki_takayama_tsuchiya-2006}, confirming the
Lin-Maldacena's gauge/gravity correspondence \cite{lin_maldacena-2006}.
They showed the following formal equivalences: the theory around each
vacuum of the supersymmetric Yang-Mills on ${\mathbb R} \times S^2$ is
equivalent to the theory around a certain vacuum of the plane wave
matrix model; the theory around each vacuum of the supersymmetric
Yang-Mills on ${\mathbb R} \times S^3$ is equivalent to the theory
around a certain vacuum of the supersymmetric Yang-Mills on ${\mathbb R}
\times S^2$ with the orbifolding condition imposed
\cite{berenstein_maldacena_nastase-2002}.
They thus made the connection between the theory around each vacuum of
the supersymmetric Yang-Mills on ${\mathbb R} \times S^3$ and the theory
around a certain vacuum of the plane wave matrix model with orbifolding
condition imposed.
In this identification, $S^3$ emerges out of a group of the concentric fuzzy
spheres.
Note that the equivalences shown in
\cite{ishiki_shimasaki_takayama_tsuchiya-2006} are classical, since the
equivalences are shown at tree level and the size of matrices are
infinite with the orbifolding condition imposed.
Recently, they extend the equivalence between the supersymmetric Yang-Mills
on ${\mathbb R} \times S^3$ and the plane wave matrix model at quantum
level \cite{ishii_ishiki_shimasaki_tsuchiya-2008}.
The equivalence is shown upto the one-loop level and the size of
matrices is finite without the orbifolding conditions.
Moreover, they derive the deconfinment phase transition of the
supersymmetric Yang-Mills on $S^1 \times S^3$ at weak coupling
region from the plane wave matrix model \cite{ishiki_kim_nishimura_tsuchiya-2008}.

In order to elucidate these proposals to construct curved spacetime in matrix models,
we investigated the effective action of the deformed IKKT matrix model
with a Myers term. Since the classical solution satisfies the commutation relation of the angular momentum,
it can be interpreted as the covariant derivatives on $S^3$
or concentric fuzzy spheres
\cite{kaneko_kitazawa_matsumoto-2007}.
In the both cases, we found that the highly divergent contributions at
the tree and one-loop level are sensitive to the UV cutoff.
However the two-loop level contributions are universal since they are
only logarithmically divergent.
We expect that the higher loop contributions are insensitive to the UV
cutoff since three-dimensional gauge theory is super renormalizable.

In the large $N$ limit, there is a well-known equivalence between a gauge theory and
a matrix model due to Eguchi and Kawai \cite{eguchi_kawai-1982}.
They proved that a large $N$ gauge theory is equivalent to a matrix
model which is dimensional reduced to zero dimension unless the
$U(1)^d$ symmetry is broken, where $d$ represents the dimension of the
original gauge theory.
However, the $U(1)^d$ symmetry is spontaneously broken in $d>2$.
So two improved versions of this large $N$ reduced model which preserve
the $U(1)^d$ symmetry was proposed.
One is the quenched reduced models \cite{bhanot_heller_neuberger-1982,
g_parisi-1982, gross_kitazawa-1982, das_wadia-1982} and the other is the
twisted reduced models \cite{gonzalezarroyo_okawa-1983a,gonzalezarroyo_okawa-1983b,
eguchi_nakayama-1983}.
However, in these models the connection is made between matrix models and
gauge theories on flat spacetime.
In this paper, we investigate the effective action of the plane wave
matrix model on a group of concentric fuzzy spheres at finite temperature.
We find that the effective action is consistent with the free
energy of the ${\cal N}=4$ supersymmetric Yang-Mills on $S^3$ in the
high temperature limit. It is because planar contributions dominates
in the high temperature limit.
We conclude that the plane wave matrix model can be used as a large $N$
reduced model to investigate
nonperturbative aspects of the ${\cal N}=4$ supersymmetric Yang-Mills theory
 on ${\mathbb R} \times S^3$.

The organization of this paper is as follows.
In section 2, we formally derive the action of the ${\cal N}=4$
supersymmetric Yang-Mills on $S^3$ from the plane wave matrix model.
In section 3, we calculate the effective action of the plane wave
matrix model around a group of concentric fuzzy spheres
at finite temperature.
Section 4 is devoted to conclusions and discussions.
Some detailed calculations are gathered in the appendix.

\section{${\cal N}=4$ supersymmetric Yang-Mills on ${\mathbb R} \times
 S^3$ as plane wave matrix model}

In this section, we formally derive the action of the supersymmetric Yang-Mills
theory on ${\mathbb R} \times S^3$ from the plane wave matrix
model in the large $N$ limit.

The authors of \cite{ishiki_shimasaki_takayama_tsuchiya-2006} observed
the following two equivalences between the vacua of different gauge theories
and the plane wave matrix model \footnote{The recent developments are
explained in the introduction.}:
\begin{description}
 \item[(i)] The supersymmetric Yang-Mills
            theory on ${\mathbb R} \times S^2$ is equivalent to the
            theory around a certain vacuum of the plane wave matrix model.
            
 \item[(i\hspace{-.1em}i)]  The
            supersymmetric Yang-Mills theory on ${\mathbb R} \times S^3$
            is equivalent to the theory around a certain vacuum of the
            supersymmetric Yang-Mills theory on ${\mathbb R} \times S^2$
            with a generalized compactification procedure in the $S^1$
            direction.
\end{description}
From the above equivalences of (i) and (i\hspace{-.1em}i), they
concluded that $S^3$ is realized by three matrices.
The three matrices is as follows:
\begin{equation}
 Y_{i}=-\mu L_{i},
\end{equation}
where
\begin{equation}
 L_i=
  \left(
   \begin{array}{ccccc}
    \rotatebox[origin=tl]{-35}
     {$
     \;\;\;\cdots\;\;\;
     \overbrace{\rotatebox[origin=c]{35}{$L_{i}^{[j_{s-1}]}$}
     \;\cdots\;
     \rotatebox[origin=c]{35}{$L_{i}^{[j_{s-1}]}$}}^{\rotatebox{35}{$n$}}
     \overbrace{\rotatebox[origin=c]{35}{$L_i^{[j_s]}$}
     \;\cdots\;
     \rotatebox[origin=c]{35}{$L_i^{[j_s]}$}}^{\rotatebox{35}{$n$}}
     \overbrace{\rotatebox[origin=c]{35}{$L_{i}^{[j_{s+1}]}$}
     \;\cdots\;
     \rotatebox[origin=c]{35}{$L_{i}^{[j_{s+1}]}$}}^{\rotatebox{35}{$n$}}
     \;\;\;\cdots\;\;\;
     $}
   \end{array}
  \right).
  \label{vacuum-pwmm}
\end{equation}
The representation matrix $L_{i}$, where  $i=1,2,3$, is a reducible
representation of $SU(2)$, and obeys the following commutation relation:
\begin{equation}
 \left[L_i,L_j\right]={\rm i}\epsilon_{ijk}L^k.
\end{equation}
$L_i^{[j_s]}$, where $s=-\infty,\cdots,\infty$, is the
$\left(2j_s+1\right) \times \left(2j_s+1\right)$ representation matrix
for the spin $j_s$ irreducible representation of $SU(2)$, and obeys the
following commutation relation:
\begin{equation}
 \left[L_i^{[j_s]},L_j^{[j_s]}\right]
  ={\rm i}\epsilon_{ijk}L^{[j_s]k}.
\end{equation}
Then, the Casimir operator of $L_i^{[j_s]}$ is that
\begin{equation}
 L_i^{[j_s]}L^{[j_s]i}
  =j_s\left(j_s+1\right)
  \mbox{\boldmath $1$}_{2j_s+1}.
\end{equation}
The matrices (\ref{vacuum-pwmm}) can be interpreted as $n$ sets of
$\infty$ fuzzy spheres with the radius $\mu\sqrt{j_s\left(j_s+1\right)}$,
where all the fuzzy spheres are concentric. 
In order to make the connection between the supersymmetric Yang-Mills
theory on ${\mathbb R} \times S^3$ and the plane wave matrix model, it
is necessary to impose the following conditions:
\begin{equation}
 j_{s}-j_{t}=\frac12(s-t),
  \hspace{5mm}
 j_{s},j_{t} \to \infty,
  \hspace{5mm}
  s,t=-\infty,\cdots,\infty.
  \label{condition-YM_PWMM}
\end{equation}

Let us start with the plane wave matrix model which is defined by the
following action:
\begin{eqnarray}
 &&S_{\rm PW}
  =\frac{1}{g_{\rm PW}^2}
  \int \frac{dt}{\mu^2}\: {\rm Tr}
  \Biggl\{
   \frac{1}{2}
   \left(D_0X_i\right)^2
   -\frac{1}{2}
   \left(
    \mu X_i
    -\frac{{\rm i}}{2}\epsilon_{ijk}
    \left[X^j,X^k\right]
   \right)^2
   \nonumber \\
 &&\hspace{25mm}
  +\frac{1}{2}\left(D_0X_m\right)^2
  -\frac{\mu^2}{8}X_m^2
  +\frac{1}{2}\left[X_i,X_m\right]^2
  +\frac{1}{4}\left[X_m,X_n\right]^2
 \nonumber \\
 &&\hspace{25mm}
  +\frac{{\rm i}}{2}\bar{\lambda}\Gamma^0D_0\lambda
  +\frac{3{\rm i}\mu}{8}\bar{\lambda}\Gamma^{123}\lambda
  -\frac{1}{2}\bar{\lambda}\Gamma^{i}
  \left[X_i,\lambda\right]
  -\frac{1}{2}\bar{\lambda}\Gamma^{m}
  \left[X_m,\lambda\right]
 \Biggr\},
 \label{PWMM}
\end{eqnarray}
where $X$ and $\lambda$ are vector and Majorana-Weyle spinor fields, and
both fields are $N \times N$ Hermitian matrices.
The vector indices $i,j,k$ and $m,n$ run over as follows: $i,j,k=1,2,3$
and $m,n=4,\cdots,9$.
The covariant derivative is given by $D_{0}{\cal O}=\partial_{0}{\cal
O}-{\rm i}\left[A_{0},{\cal O}\right]$.
The radius of $S^3$ is fixed to $2/\mu$.

Let us consider such a large $N$ limit as follows:
\begin{equation}
 X_{i}(t)
  \to
  -\mu\nabla_{i}+B_{i}(t,\mbox{\boldmath $x$}),
  \hspace{5mm}
  X_{m}(t)
  \to
  X_{m}(t,\mbox{\boldmath $x$}),
  \hspace{5mm}
  \lambda(t)
  \to
  \lambda(t,\mbox{\boldmath $x$}),
\end{equation}
where $\nabla_{i}$ and $B_{i}$ are derivatives and space components of
gauge fields on $S^3$ that are defined by Killing vectors (See
ref. \cite{kaneko_kitazawa_matsumoto-2007} for a review on this subject):
\begin{equation}
 \nabla_{i}=K_{i}^{a}\partial_{a},
  \hspace{10mm}
  B_{i}(t,\mbox{\boldmath $x$})
  =K_{i}^{a}A_{a}(t,\mbox{\boldmath $x$}),
\end{equation}
where $a=\theta,\phi,\psi$.
The non-vanishing components of Killing vectors are given by
\begin{equation}
 K_{1}^{\theta}=\mu,
  \hspace{5mm}
  K_{2}^{\phi}=\frac{\mu}{\sin\theta},
  \hspace{5mm}
  K_{2}^{\psi}=-\frac{\mu\cos\theta}{\sin\theta},
  \hspace{5mm}
  K_{3}^{\psi}=1.
\end{equation}
For example, we consider the following term in the action of the plane
wave matrix model:
\begin{equation}
 \frac{1}{g_{\rm PW}^2}
  \int\!\frac{dt}{\mu^2}{\rm Tr}
  \left\{
   -\frac12
   \left(
    \mu X_{i}
    -\frac{{\rm i}}{2}\epsilon_{ijk}\left[X^{j},X^{k}\right]
   \right)^2
  \right\}.
 \label{bosonic_term}
\end{equation}
By taking the large $N$ limit, the term (\ref{bosonic_term}) can be
rewritten as follows:
\begin{eqnarray}
 &&\frac{1}{g_{\rm PW}^2}
  \int\!\frac{dt}{\mu^2}{\rm Tr}
  \Biggl\{
 -\frac12
  \biggl(
    -\mu^2\nabla_{i}
    +\mu B_{i}
  \nonumber \\
 &&\hspace{20mm}
     -\frac{{\rm i}}{2}\epsilon_{ijk}
    \left(
     \mu^2\left[\nabla^{j},\nabla^{k}\right]
     -\mu
     \left(\nabla^{j}B^{k}-\nabla^{k}B^{j}\right)
     +\left[B^{j},B^{k}\right]
    \right)
  \biggr)^2
  \Biggr\}.
\end{eqnarray}
From the commutation relation for the derivatives on
$S^3$:
\begin{equation}
 \left[\nabla_{i},\nabla_{j}\right]
  ={\rm i}\epsilon_{ijk}\nabla^{k},
  \label{derivative_rel}
\end{equation}
we can obtain the following relation:
\begin{equation}
 K_{i}^{a}\partial_{a}K_{j}^{b}
  -K_{j}^{a}\partial_{a}K_{i}^{b}
  ={\rm i}\epsilon_{ijk}K_{k}^{b}.
  \label{killing_rel}
\end{equation}
Then, we can get the following equation by using (\ref{derivative_rel}) and
(\ref{killing_rel}):
\begin{eqnarray}
 &&\frac{1}{g_{\rm PW}^2}
  \int\!\frac{dt}{\mu^2}{\rm Tr}
  \left\{
   -\frac12
   \left(
    \frac{{\rm i}\mu}{2}
    \epsilon_{ijk}K_{a}^{j}K_{b}^{k}
    \left(
     \partial^{a}A^{b}-\partial^{b}A^{a}
    \right)
    -\frac{{\rm i}}{2}
    \epsilon_{ijk}K_{a}^{j}K_{b}^{k}
    \left[A^{a},A^{b}\right]
   \right)^2
  \right\}
  \nonumber \\
 &&\hspace{10mm}
  =\frac{\mu N}{16\pi^2 g_{\rm PW}^2n}
  \int\!d^4x\sqrt{g} \, {\rm tr}
  \left\{
   -\frac14g_{ac}g_{bd}F^{ab}F^{cd}
  \right\},
\end{eqnarray}
where
\begin{equation}
 F_{ab}
  =\partial_{a}A_{b}
  -\partial_{b}A_{a}
  -{\rm i}\left[A_{a},A_{b}\right].
\end{equation}
Note that we also rescaled the derivatives on $S^3$ as follows: 
\begin{equation}
 {\rm i}\mu\partial_{a}
  \to
  \partial_{a}.
\end{equation}
It is because we have the following correspondence in the large $N$ limit:
\begin{equation}
 {\rm Tr}
  \rightarrow
  \frac{N}{{\rm Vol}(S^3)n}
  \int\!d^3x \sqrt{g} \, {\rm tr}
\end{equation}
where $tr$ denotes the trace operation over $SU(n)$ gauge group.
Similarly, taking the large $N$ limit of the other terms in the
action of the plane wave matrix model, we can obtain the action of
supersymmetric Yang-Mills theory on ${\mathbb R} \times S^3$ as follows: 
\begin{eqnarray}
 &&S_{\rm SYM}
  =\frac{2}{g_{\rm SYM}^2n}
  \int d^4x\sqrt{g}\,
  {\rm tr}
  \Biggl\{
  -\frac{1}{4}F_{\mu\nu}F^{\mu\nu}
  +\frac{1}{2}D_{\mu}X_{m}D^{\mu}X_{m}
  -\frac{1}{12}RX_{m}^2
  \Biggr.
  \nonumber \\
 &&\hspace{4cm}\;\;\
  \Biggl.
   +\frac{{\rm i}}{2}\bar{\lambda}\Gamma^{\mu}D_{\mu}\lambda
   -\frac{1}{2}\bar{\lambda}\Gamma^m[X_{m},\lambda]
   +\frac{1}{4}[X_{m},X_{n}]^2
  \Biggr\},
 \label{SYM}
\end{eqnarray}
where $\mu=0,1,2,3$, and $R$ is the scalar curvature of $S^3$.

\section{Effective action for plane wave matrix model}

In the preceding section, we have summarized formal arguments for 
the equivalence between the gauge theory on $R \times S^3$ and a certain vacuum
configuration of the plane wave matrix model.
However they are formal in the sense that they need to consider the large $N$ limit.
Therefore their validity is not automatic especially at the nonperturbative level, since
we need to work with finite $N$. In this section we work with finite $N$, namely finite
size matrices.  To be precise, we introduce the two cutoffs in the theory
with respect to the size and number of the concentric fuzzy spheres.
We also put $n=1$.
In such a set up, we investigate the effective action perturbatively
to check to what extent formal arguments can be justified.

\subsection{One-loop effective action at zero temperature}

In this subsection, we evaluate the one-loop effective action of the
plane wave matrix model around $S^3$ background at zero temperature.

As in the ordinary background field method in quantum field theories, we
decompose matrices $X$ and $\lambda$ into the backgrounds and quantum
fluctuations, respectively as follows:
\begin{eqnarray}
 &&X_{i}=p_{i}+x_{i},
  \hspace{10mm}
 X_{m}=p_{m}+x_{m},
 \nonumber \\
 &&\lambda=\chi+\varphi,
  \label{decom_matrices}
\end{eqnarray}
where $p_{i}$, $p_{m}$ and $\chi$ are backgrounds, and $x_{i}$, $x_{m}$
and $\varphi$ are quantum fluctuations.
Then, we substitute the decomposed matrices (\ref{decom_matrices}) into
the action of the plane wave matrix model, and expand around backgrounds
up to the forth order with respect to quantum fluctuations.
The expanded action is expressed as follows:
\begin{equation}
 S_{\rm PW}
  =S_{\rm PW}^{(0)}
  +S_{\rm PW}^{(1)}
  +S_{\rm PW}^{(2)}
  +S_{\rm PW}^{(3)}
  +S_{\rm PW}^{(4)},
\end{equation}
where
\begin{eqnarray}
 &&S_{\rm PW}^{(0)}
  =\frac{1}{g_{\rm PW}^2\mu^2}
  \int\!dt\,
  {\rm Tr}
  \Biggl\{
   \frac{1}{2}
   \bigl(\partial_{0}p_{i}\bigr)^2
   -\frac12\mu^2p_{i}^2
   +\frac{{\rm i}}{2}\epsilon_{ijk}
   p^{i}\bigl[p^{j},p^{k}\bigr]
   +\frac14
   \bigl[p_{i},p_{j}\bigr]^2
   \nonumber \\
 &&\hspace{30mm}
  +\frac{1}{2}
  \bigl(\partial_{0}p_{m}\bigr)^2
  -\frac{\mu^2}{8}p_{m}^2
  +\frac12
  \bigl[p_{i},p_{m}\bigr]^2
  +\frac14
  \bigl[p_{m},p_{n}\bigr]^2
  \nonumber \\
 &&\hspace{30mm}
  +\frac{{\rm i}}{2}
  \bar{\chi}\Gamma^{0}\partial_{0}\chi
  +\frac{3{\rm i}\mu}{8}
  \bar{\chi}\Gamma^{123}\chi
  -\frac12
  \bar{\chi}\Gamma^{i}\bigl[p_{i},\chi\bigr]
  -\frac12
  \bar{\chi}\Gamma^{m}\bigl[p_{m},\chi\bigr]
  \Biggr\},
 \\
 \nonumber \\
 &&S_{\rm PW}^{(1)}
  =\frac{1}{g_{\rm PW}^2\mu^2}
  \int\!dt\,
  {\rm Tr}
  \Biggl\{
   -x_{k}
   \biggl(
    \partial_{0}^2p^{k}
    +\mu^2p^{k}
    +\frac{3{\rm i}}{2}\mu\epsilon^{ijk}
    \bigl[p_{i},p_{j}\bigr]
    \nonumber \\
 &&\hspace{70mm}
  +\bigl[
    p_{i},[p^{i},p^{k}]
   \bigr]
  +\bigl[
    p_{m},[p^{m},p^{k}]
   \bigr]
  -\frac12
  \bigl\{\bar{\chi}\Gamma^{k},\chi\bigr\}
  \biggr)
   \nonumber \\
 &&\hspace{30mm}
  -x_{n}
  \biggl(
   \partial_{0}^2p^{n}
   +\frac{\mu^2}{4}p^{n}
   +\bigl[p_{i},[p^{i},p^{n}]\bigr]
   +\bigl[p_{m},[p^{m},p^{n}]\bigr]
   -\frac12
   \bigl\{\bar{\chi}\Gamma^{n},\chi\bigr\}
  \biggr)
  \nonumber \\
 &&\hspace{30mm}
  -{\rm i}\bigl(\partial_{0}p_{i}\bigr)
  \bigl[A_{0},p^{i}\bigr]
  -{\rm i}\bigl(\partial_{0}p_{m}\bigr)
  \bigl[A_{0},p^{m}\bigr]
  -\frac12
  \bar{\chi}\Gamma^{0}\bigl[A_{0},\chi\bigr]
  \nonumber \\
 &&\hspace{30mm}
  +\bar{\varphi}
  \biggl(
   \frac{{\rm i}}{2}
   \Gamma^{0}\partial_{0}\chi
   +\frac{3{\rm i}\mu}{8}
   \Gamma^{123}\chi
   -\frac12
   \Gamma^{i}\bigl[p_{i},\chi\bigr]
   -\frac12
   \Gamma^{m}\bigl[p_{m},\chi\bigr]
  \biggr)
  \nonumber \\
 &&\hspace{30mm}
  +\biggl(
    \frac{{\rm i}}{2}
    \bigl(\partial_{0}\bar{\chi}\bigr)\Gamma^{0}
    +\frac{3{\rm i}\mu}{8}
    \bar{\chi}\Gamma^{123}
    -\frac12
    \bigl[\bar{\chi}\Gamma^{i},p_{i}\bigr]
    -\frac12
    \bigl[\bar{\chi}\Gamma^{m},p_{m}\bigr]
   \biggr)
  \varphi
 \Biggr\},
 \\
 \nonumber \\
 &&S_{\rm PW}^{(2)}
  =\frac{1}{g_{\rm PW}^2\mu^2}
  \int\!dt\,
  {\rm Tr}
  \Biggl\{
   \frac{1}{2}
   \bigl(\partial_{0}x_{i}\bigr)^2
   -{\rm i}\bigl(\partial_{0}p_{i}\bigr)
   \bigl[A_{0},x^{i}\bigr]
   -{\rm i}\bigl(\partial_{0}x_{i}\bigr)
   \bigl[A_{0},p^{i}\bigr]
   -\frac12
   \bigl[A_{0},p_{i}\bigr]^2
   \nonumber \\
 &&\hspace{30mm}
  -\frac12\mu^2x_{i}^2
   -\frac{3{\rm i}}{2}\mu\epsilon^{ijk}
   x_{i}\bigl[p_{k},x_{j}\bigr]
   +\frac12
   \bigl[p_{i},x_{j}\bigr]^2
   -\frac12
   \bigl[p_{i},x^{i}\bigr]^2
   +\bigl[p_{i},p_{j}\bigr]
   \bigl[x^{i},x^{j}\bigr]
   \nonumber \\
 &&\hspace{30mm}
  +\frac{1}{2}
  \bigl(\partial_{0}x_{m}\bigr)^2
  -{\rm i}\bigl(\partial_{0}p_{m}\bigr)
  \bigl[A_{0},x^{m}\bigr]
  -{\rm i}\bigl(\partial_{0}x_{m}\bigr)
  \bigl[A_{0},p^{m}\bigr]
  -\frac12
  \bigl[A_{0},p_{m}\bigr]^2
  \nonumber \\
 &&\hspace{30mm}
  -\frac{\mu^2}{8}x_{m}^2
  +\frac12
  \bigl[p_{i},x_{m}\bigr]^2
  +\frac12
  \bigl[p_{m},x_{i}\bigr]^2
  +2\bigl[p_{i},p_{m}\bigr]
  \bigl[x^{i},x^{m}\bigr]
  \nonumber \\
 &&\hspace{30mm}
  -\bigl[p_{i},x^{i}\bigr]
  \bigl[p_{m},x^{m}\bigr]
  +\frac12
  \bigl[p_{m},x_{n}\bigr]^2
  -\frac12
  \bigl[p_{m},x^{m}\bigr]^2
  +\bigl[p_{m},p_{n}\bigr]
  \bigl[x^{m},x^{n}\bigr]
  \nonumber \\
 &&\hspace{30mm}
  +\frac{{\rm i}}{2}
  \bar{\varphi}\Gamma^{0}\partial_{0}\varphi
  +\frac12
  \bar{\chi}\Gamma^{0}\bigl[A_{0},\varphi\bigr]
  +\frac12
  \bar{\varphi}\Gamma^{0}\bigl[A_{0},\chi\bigr]
  +\frac{3{\rm i}\mu}{8}
  \bar{\varphi}\Gamma^{123}\varphi
  \nonumber \\
 &&\hspace{30mm}
  -\frac12
  \bar{\chi}\Gamma^{i}\bigl[x_{i},\varphi\bigr]
  -\frac12
  \bar{\varphi}\Gamma^{i}\bigl[p_{i},\varphi\bigr]
  -\frac12
  \bar{\varphi}\Gamma^{i}\bigl[x_{i},\chi\bigr]
  \nonumber \\
 &&\hspace{30mm}
  -\frac12
  \bar{\chi}\Gamma^{m}\bigl[x_{m},\varphi\bigr]
  -\frac12
  \bar{\varphi}\Gamma^{m}\bigl[p_{m},\varphi\bigr]
  -\frac12
  \bar{\varphi}\Gamma^{m}\bigl[x_{m},\chi\bigr]
  \Biggr\},
  \\
  \nonumber \\
 &&S_{\rm PW}^{(3)}
  =\frac{1}{g_{\rm PW}^2\mu^2}
  \int\!dt\,
  {\rm Tr}
  \Biggl\{
  -{\rm i}\bigl(\partial_{0}x_{i}\bigr)
  \bigl[A_{0},x^{i}\bigr]
  -\bigl[A_{0},p_{i}\bigr]
  \bigl[A_{0},x^{i}\bigr]
  +\frac{{\rm i}}{2}\mu\epsilon^{ijk}
  x_{i}\bigl[x_{j},x_{k}\bigr]
   \nonumber \\
 &&\hspace{30mm}
  +\bigl[p_{i},x_{j}\bigr]
  \bigl[x^{i},x^{j}\bigr]
  -{\rm i}\bigl(\partial_{0}x_{m}\bigr)
  \bigl[A_{0},x^{m}\bigr]
  -\bigl[A_{0},p_{m}\bigr]
  \bigl[A_{0},x^{m}\bigr]
  \nonumber \\
 &&\hspace{30mm}
  +\bigl[p_{i},x_{m}\bigr]
  \bigl[x^{i},x^{m}\bigr]
  +\bigl[p_{m},x_{i}\bigr]
  \bigl[x^{m},x^{i}\bigr]
  +\bigl[p_{m},x_{n}\bigr]
  \bigl[x^{m},x^{n}\bigr]
   \nonumber \\
 &&\hspace{30mm}
  +\frac12
  \bar{\varphi}\Gamma^{0}\bigl[A_{0},\varphi\bigr]
  -\frac12
  \bar{\varphi}\Gamma^{i}\bigl[x_{i},\varphi\bigr]
  -\frac12
  \bar{\varphi}\Gamma^{m}\bigl[x_{m},\varphi\bigr]
  \Biggr\},
  \\
  \nonumber \\
 &&S_{\rm PW}^{(4)}
  =\frac{1}{g_{\rm PW}^2\mu^2}
  \int\!dt\,
  {\rm Tr}
  \Biggl\{
  -\frac12
  \bigl[A_{0},x_{i}\bigr]^2
  +\frac14
  \bigl[x_{i},x_{j}\bigr]^2
  \nonumber \\
 &&\hspace{60mm}
  -\frac12
  \bigl[A_{0},x_{m}\bigr]^2
  +\frac12
  \bigl[x_{i},x_{m}\bigr]^2
  +\frac14
  \bigl[x_{m},x_{n}\bigr]^2
  \Biggr\}.
\end{eqnarray}
Since we need to fix the gauge invariance in the action, we add the
gauge fixing and the Faddeev-Popov terms:
\begin{eqnarray}
 &&S_{\rm GF}
  =\frac{1}{g_{\rm PW}^2\mu^2}
  \int\!dt\,
  {\rm Tr}
  \Biggl\{
   -\frac12
   \left(
    \partial_{0}A_{0}
    +{\rm i}\left[p_{i},X^{i}\right]
    +{\rm i}\left[p_{m},X^{m}\right]
   \right)^2
  \Biggr\},
  \\
 &&S_{\rm FP}
  =\frac{1}{g_{\rm PW}^2\mu^2}
  \int\!dt\,
  {\rm Tr}
  \Biggl\{
   -b\partial_{0}D_{0}c
   -b\left[
     p_{i},\left[X^{i},c\right]
    \right]
   -b\left[
     p_{m},\left[X^{m},c\right]
    \right]
  \Biggr\},
  \label{GF-FP}
\end{eqnarray}
where $c$ and $b$ are ghost and anti-ghost fields, respectively.

We substitute the matrices $Y_i$ which are the classical solution
of the plane wave matrix model for backgrounds as follows:
\begin{eqnarray}
 &&p_i=Y_i=-\mu L_{i},
  \hspace{10mm}
  p_{m}=0,
  \nonumber \\
 &&\chi=0,
  \label{back_grounds}
\end{eqnarray}
where
\begin{equation}
 L_i=
  \left(
   \begin{array}{ccccc}
    \rotatebox[origin=tl]{-35}
     {$
     \rotatebox[origin=c]{35}{$L_{i}^{[j_{1}]}$}
     \;\;\;\cdots\;\;\;
     \rotatebox[origin=c]{35}{$L_{i}^{[j_{s}]}$}
     \;\;\;\cdots\;\;\;
     \rotatebox[origin=c]{35}{$L_i^{[j_{2\Lambda}]}$}
     $}
   \end{array}
  \right).
\end{equation}
Here, we introduce a cutoff on $s$ at $2\Lambda$ and on the matrix
size of $L_{i}^{[j_{s}]}$ at $2j_{s}+1=N_{0}+s$, and the matrix size
${N}$ of $L_{i}$ is finite as follows:
\begin{equation}
 {N}
 =\left(2j_{1}+1\right)
 +\cdots
 +\left(2j_{s}+1\right)
 +\cdots
 +\left(2j_{2\Lambda}+1\right).
\end{equation}
Then, we can obtain the following action:
\begin{eqnarray}
 \tilde{S}_{\rm PW}
  &=&S_{\rm PW}+S_{\rm GF}+S_{\rm FP}
  \nonumber \\
 &=&
  \frac{1}{g_{\rm PW}^2\mu^2}
  \int\!dt\!
  \sum_{s,t}{\rm Tr}
  \nonumber \\
 &&\times
  \Biggl\{
   -\frac12 x_{i}^{(s,t)}
   \biggl(
    -\delta^{ij}\partial_{0}^2
    +\delta^{ij}\mu^2 {\cal L}_{i}^2
    +2\mu^2
    \left[{\cal L}_{i},{\cal L}_{j}\right]
    +\mu^2\delta^{ij}
    -3{\rm i}\mu^2\epsilon^{ijk}{\cal L}_{k}
   \biggr)
   x_{j}^{(t,s)}
   \nonumber \\
 &&\hspace{5mm}
  -\frac12x_{m}^{(s,t)}
  \biggl(
   -\delta^{mn}\partial_{0}^2
   +\delta^{mn}\mu^2 {\cal L}_{i}^2
   +\frac{\mu^2}{4}\delta^{mn}
  \biggr)
  x_{n}^{(t,s)}
  \nonumber \\
 &&\hspace{5mm}
  -\frac12 A_{0}^{(s,t)}
  \biggl(
   -\partial_{0}^2
   +\mu^2 {\cal L}_{i}^2
  \biggr)
  A_{0}^{(t,s)}
  -b^{(s,t)}
  \biggl(
   -\partial_{0}^2
   +\mu^2 {\cal L}_{i}^2
  \biggr)
  c^{(t,s)}
  \nonumber \\
 &&\hspace{5mm}
  -\frac12 \bar{\varphi}^{(s,t)}
  \biggl(
   -{\rm i}\Gamma^{0}\partial_{0}
   -\mu\Gamma^{i} {\cal L}_{i}
   -\frac{3{\rm i}\mu}{4}\Gamma^{123}
  \biggr)
  \varphi^{(t,s)}
  \nonumber \\
 &&\hspace{5mm}
  -{\rm i}\left(\partial_{0}x_{i}^{(s,t)}\right)
   \left[A_{0},x^{i}\right]^{(t,s)}
   +\mu {\cal L}_{i}A_{0}^{(s,t)}
   \left[A_{0},x^{i}\right]^{(t,s)}
   +\frac{{\rm i}}{2}\mu
   \epsilon^{ijk}x_{i}^{(s,t)}
   \left[x_{j},x_{k}\right]^{(t,s)}
  \nonumber \\
 &&\hspace{5mm}
  -\mu {\cal L}_{i} x_{j}^{(s,t)}
  \left[x^{i},x^{j}\right]^{(t,s)}
  -{\rm i}\left(\partial_{0}x_{m}^{(s,t)}\right)
  \left[A_{0},x^{m}\right]^{(t,s)}
  -\mu {\cal L}_{i} x_{m}^{(s,t)}
  \left[x^{i},x^{m}\right]^{(t,s)}
  \nonumber \\
 &&\hspace{5mm}
  +{\rm i}b^{(s,t)}\partial_{0}
  \left[A_{0},c\right]^{(t,s)}
  +\mu b^{(s,t)} {\cal L}_{i}
  \left[x^{i},c\right]^{(t,s)}
  +\frac12\bar{\varphi}^{(s,t)}\Gamma^{0}
  \left[A_{0},\varphi\right]^{(t,s)}
  \nonumber \\
 &&\hspace{5mm}
  +\frac12\bar{\varphi}^{(s,t)}\Gamma^{i}
  \left[x_{i},\varphi\right]^{(t,s)}
  +\frac12\bar{\varphi}^{(s,t)}\Gamma^{m}
  \left[x_{m},\varphi\right]^{(t,s)}
  +\frac12
  \left[A_{0},x_{i}\right]^{(s,t)2}
  \nonumber \\
 &&\hspace{5mm}
  +\frac14
  \left[x_{i},x_{j}\right]^{(s,t)2}
  +\frac12
  \left[A_{0},x_{m}\right]^{(s,t)2}
   +\frac12
   \left[x_{i},x_{m}\right]^{(s,t)2}
   +\frac14
   \left[x_{m},x_{n}\right]^{(s,t)2}
  \Biggr\},
  \nonumber \\
  \label{Euclidean_S-PW}
\end{eqnarray}
where the suffix $(s,t)$ represents the $(s,t)$ block in the
${N}\times{N}$ matrix.
We introduce the following operation:
\begin{equation}
 {\cal L}_{i} M
 =\left[L_{i},M\right],
\end{equation}
which act on a $\left(2j_{s}+1\right) \times \left(2j_{t}+1\right)$
matrix $M$. 
Note that the above action is obtained after the Wick rotation for the time components
as follows:
\begin{equation}
 t \to {\rm i}t,
  \hspace{5mm}
  A_{0} \to {\rm i}A_{0},
  \hspace{5mm}
  \Gamma^{0} \to {\rm i}\Gamma^{0}.
\end{equation}

The classical action: $S_{\rm PW}^{(0)}$ vanishes for the backgrounds we
consider here, and the first order action with respect to the quantum
fluctuations: $S_{\rm PW}^{(1)}$ also vanishes, because the backgrounds
satisfy the equations of motion for the plane wave matrix model.
We need the quadratic action with respect to the quantum
fluctuations $S_{\rm PW}^{(2)}$ to calculate an one-loop effective
action of the plane wave matrix model.
To simplify the following calculations, we introduce the notations as
follows:
\begin{equation}
 p_{0}=-{\rm i}\partial_{0},
  \hspace{10mm}
  p_{i}=-\mu L_{i},
  \hspace{10mm}
  p_{m}=0.
\end{equation}
Then, the action is given by
\begin{eqnarray}
 &&\tilde{S}_{\rm PW}^{\rm 1-loop}
  =\frac{2\pi N_{0}}{g_{\rm PW}^2\mu^2}
  \sum_{s,t}
  \sum_{J,M}
  \nonumber \\
 &&\hspace{20mm}
  \times
  \Biggl[
  \sum_{l}
  \Biggl\{
   -\frac12 x_{ilJM}^{(s,t)}
   \biggl(
    \delta^{ij}{\cal P}_{{\rm B}}^2
    +2{\rm i}{\cal F}_{{\rm B}}^{ij}
    +\mu^2\delta^{ij}
    +3{\rm i}\mu\epsilon^{ijk}{\cal P}_{{\rm B}k}
  \biggr)
  x_{jlJM}^{(s,t)\dagger}
  \nonumber \\
 &&\hspace{35mm}
  -\frac12x_{mlJM}^{(s,t)}
  \biggl(
   \delta^{mn}{\cal P}_{{\rm B}}^2
   +\frac{\mu^2}{4}\delta^{mn}
  \biggr)
  x_{nlJM}^{(s,t)\dagger}
  \nonumber \\
 &&\hspace{35mm}
  -\frac12 A_{0lJM}^{(s,t)}
  \biggl(
   {\cal P}_{{\rm B}}^2
  \biggr)
  A_{0lJM}^{(s,t)\dagger}
  -b_{lJM}^{(s,t)}
  \biggl(
   {\cal P}_{{\rm B}}^2
  \biggr)
  c_{lJM}^{(s,t)\dagger}
  \Biggr\}
  \nonumber \\
 &&\hspace{25mm}
  +\sum_{h}
  \Biggl\{
  -\frac12
  \bar{\varphi}_{hJM}^{(s,t)}
  \biggl(
   \Gamma^{\mu}{\cal P}_{{\rm F}\mu}
   +\frac{3{\rm i}\mu}{4}\Gamma^{123}
  \biggr)
  \varphi_{hJM}^{(s,t)\dagger}
  \Biggr\}
  \Biggr],
\end{eqnarray}
where
\begin{equation}
 {\cal P}_{i}M
  =\bigl[p_{i},M\bigr],
  \hspace{5mm}
  {\cal F}_{ij}M
  =\bigl[f_{ij},M\bigr],
  \hspace{5mm}
  f_{ij}
  =-{\rm i}\bigl[p_{i},p_{j}\bigr],
\end{equation}
and the index $\mu$ runs from $0$ to $3$.

By using the above action $\tilde{S}_{\rm PW}^{\rm 1-loop}$, we can
calculate the one-loop effective action of the plane wave matrix model
on $S^3$ as follows:
\begin{equation}
 W=-\log\int\!
  dx_{i}\,dx_{m}\,dA_{0}\,db\,dc\,d\varphi\,
  {\rm e}^{-\tilde{S}_{\rm PW}^{\rm 1-loop}}.
\end{equation}
First, we evaluate the bosonic parts of the effective action as follows:
\begin{eqnarray}
 &&W_{\rm B}=
  \sum_{s,t}
  \sum_{l}
  \Biggl\{
   \frac12{\rm Tr}\log
  \biggl(
   \delta^{ij}{\cal P}_{{\rm B}}^2
   +2{\rm i}{\cal F}_{{\rm B}}^{ij}
   +\mu^2\delta^{ij}
   +3{\rm i}\mu\epsilon^{ijk}{\cal P}_{{\rm B}k}
  \biggr)
  \nonumber \\
 &&\hspace{10mm}
  +\frac12{\rm Tr}\log
  \biggl(
   \delta^{mn}{\cal P}_{{\rm B}}^2
   +\frac{\mu^2}{4}\delta^{mn}
  \biggr)
  +\frac12{\rm Tr}\log
  \biggl(
   {\cal P}_{{\rm B}}^2
  \biggr)
  -{\rm Tr}\log
  \biggl(
   {\cal P}_{{\rm B}}^2
  \biggr)
  \Biggr\}.
  \label{bosonic-eff_action}
\end{eqnarray}
We expand the bosonic parts (\ref{bosonic-eff_action}) of the effective
action into the inverse power series of ${\cal P}_{{\rm
B}}^2=\left(\omega_{l}^2+\mu^2J\left(J+1\right)\right)$.
In this way, we obtain the leading term of the bosonic parts of the
one-loop effective action as follows:
\begin{eqnarray}
 &&W_{\rm B}\sim
  \sum_{s,t}
  \sum_{l}
  \left\{
   4{\rm Tr}\log
   \biggl({\cal P}_{{\rm B}}^2\biggr)
  +\frac{9\mu^2}{4}
  {\rm Tr}
   \frac{1}{{\cal P}_{{\rm B}}^2}
   -\frac{\mu^2}{2}
   {\rm Tr}{\cal P}_{{\rm B}i}^2
   \Biggl(
    \frac{1}{{\cal P}_{{\rm B}}^2}
   \Biggr)^2
  \right\},
\end{eqnarray}
where ${\cal P}_{{\rm
Bi}}^2=\mu^2J\left(J+1\right)$.
Then, we evaluate the fermionic parts of the effective action as
follows:
\begin{eqnarray}
 &&W_{\rm F}=
  -\sum_{s,t}
  \sum_{h}
  \frac14{\rm Tr}\log
  \left(
   {\cal P}_{{\rm F}}^2
   +\frac{{\rm i}}{2}
   \Gamma^{\mu\nu}{\cal F}_{{\rm F}\mu\nu}
   -\frac{{\rm 3i}}{4}
   \Gamma^{\mu\nu}{\cal F}_{{\rm F}\mu\nu}
   +\frac{9\mu^2}{16}
  \right).
  \label{fermionic-eff_action}
\end{eqnarray}
Similarly, we expand the fermionic parts (\ref{fermionic-eff_action}) of
the effective action into the power series of
${\cal P}_{{\rm F}}^2=\left(\omega_{h}^2+\mu^2J\left(J+1\right)\right)$.
So, we obtain the leading term of the fermionic parts of the
one-loop effective action as follows:
\begin{eqnarray}
 &&W_{\rm F}\sim
  \sum_{s,t}
  \sum_{h}
  \left\{
   -4{\rm Tr}\log
   \biggl({\cal P}_{{\rm F}}^2\biggr)
   -\frac{9\mu^2}{4}
   {\rm Tr}
   \frac{1}{{\cal P}_{{\rm F}}^2}
   +\frac{\mu^2}{2}
   {\rm Tr}
   {\cal P}_{{\rm F}i}^2
   \Biggl(
    \frac{1}{{\cal P}_{{\rm F}}^2}
   \Biggr)^2
  \right\}.
\end{eqnarray}
where ${\cal P}_{{\rm
Fi}}^2=\mu^2J\left(J+1\right)$.
Therefore, we find that the one-loop effective action of the plane wave
matrix model on $S^3$ vanishes to the next leading order in this expansion:
\begin{equation}
 W=W_{\rm B}+W_{\rm F}
  \sim 0.
\end{equation}
In fact, it vanishes exactly due to supersymmetry \cite{shin_yoshida-2004}.

\subsection{Free energy of ${\cal N}=4$ supersymmetric Yang-Mills on $S^3$}

In this subsection, we calculate the effective action of the plane wave
matrix model on $S^3$ at finite temperature up to two-loop level.
In order to study the plane wave matrix model on $S^3$ at finite
temperature, we compactify the Eucliedean time direction with a
periodicity $\beta=1/T$, where $T$ is a temperature.
Thus, we impose the constraint of periodicity for the bosonic, ghost
and anti-ghost fields as follows: 
\begin{eqnarray}
 &&x_{i}^{(s,t)}(0)=x_{i}^{(s,t)}(\beta),
  \hspace{5mm}
  x_{m}^{(s,t)}(0)=x_{m}^{(s,t)}(\beta),
  \hspace{5mm}
  A_{0}^{(s,t)}(0)=A_{0}^{(s,t)}(\beta),
  \nonumber \\
 &&c^{(s,t)}(0)=c^{(s,t)}(\beta),
  \hspace{5mm}
  b^{(s,t)}(0)=b^{(s,t)}(\beta).
\end{eqnarray}
So, we can obtain the conditions for frequencies in
(\ref{bosonic-eff_action}) as follows:
\begin{equation}
 \omega_{l}=2\pi lT,
\end{equation}
where $l$ is the integer.
On the other hand, we impose the constraint of anti-periodicity for the
fermion fields as follows:
\begin{equation}
 \varphi^{(s,t)}\left(0\right)
  =-\varphi^{(s,t)}\left(\beta\right),
\end{equation}
and hence
\begin{equation}
 \omega_{h}=2\pi hT,
\end{equation}
where $h$ is the half-integers.

Therefore, we can obtain the one-loop effective action of the plane wave
matrix model on $S^3$ at finite temperature as follows:
\begin{eqnarray}
 &&\hat{W}^{\rm 1-loop}
  =\sum_{s,t}
 \Biggl\{
  \sum_{l}
  4{\rm Tr}\log
  \biggl(
   {\cal P}_{{\rm B}}^2
  \biggr)
  -\sum_{h}
  4{\rm Tr}\log
  \biggl(
   {\cal P}_{{\rm F}}^2
  \biggr)
  \nonumber \\
 &&\hspace{30mm}
  +\sum_{l}
   \frac{9\mu^2}{4}
   {\rm Tr}
   \frac{1}{{\cal P}_{{\rm B}}^2}
   -\sum_{h}
   \frac{9\mu^2}{4}
   {\rm Tr}
   \frac{1}{{\cal P}_{{\rm F}}^2}
  \nonumber \\
 &&\hspace{40mm}
  -\sum_{l}
   \frac{\mu^2}{2}
   {\rm Tr}
   {\cal P}_{{\rm B}i}^2
   \Biggl(
    \frac{1}{{\cal P}_{{\rm B}}^2}
   \Biggr)^2
   +\sum_{h}
   \frac{\mu^2}{2}
   {\rm Tr}
   {\cal P}_{{\rm F}i}^2
   \Biggl(
    \frac{1}{{\cal P}_{{\rm F}}^2}
   \Biggr)^2
  \Biggr\}.
\end{eqnarray}
However, since the supersymmetry is broken at finite temperature, the
contributions from the bosons and fermions do not cancel each other.
For example, we consider the leading terms of the one-loop effective
action as follows:
\begin{eqnarray}
 &&\hat{W}_{(0)}^{\rm 1-loop}
  =\sum_{s,t}
  \left\{
   \sum_{l}
   4{\rm Tr}\log
   \biggl(
    \left(2\pi lT\right)^2
    +\mu^2J\left(J+1\right)
   \biggr)
   \right.
  \nonumber \\
 &&\hspace{40mm}
  \left.
    -\sum_{h}
   4{\rm Tr}\log
   \biggl(
    \left(2\pi hT\right)^2
    +\mu^2J\left(J+1\right)
   \biggr)
   \right\}.
\end{eqnarray}
It is easy to calculate the sums over $l$ and $h$ by using the following
formulae:
\begin{eqnarray}
 &&\sum_{l=-\infty}^{\infty}
  \log\left(l^2\pi^2+z^2\right)
  -\sum_{l=1}^{\infty}
  2\log\left(l^2\pi^2\right)
  =2\log\sinh z,
  \\
 &&\sum_{h=-\infty}^{\infty}
  \log\left(h^2\pi^2+z^2\right)
  -\sum_{h=1/2}^{\infty}
  2\log\left(h^2\pi^2\right)
  =2\log\cosh z,
\end{eqnarray}
where $z$ is the complex number.
We may discard the infinite constants
which do not depend on physical parameters.
Thus, we can obtain the leading terms of the one-loop effective action
as follows:
\begin{eqnarray}
 \hat{W}_{(0)}^{\rm 1-loop}
  &=&\sum_{s,t}
  \left\{
   8{\rm Tr}\log\sinh
   \left(
    \frac{\mu\sqrt{J\left(J+1\right)}}{2T}
   \right)
   -8{\rm Tr}\log\cosh
   \left(
    \frac{\mu\sqrt{J\left(J+1\right)}}{2T}
   \right)
  \right\}
  \nonumber \\
 &=&\sum_{s,t}
  \sum_{J,M}
  8\log
  \left(
   \frac{\exp\left(\mu\sqrt{J\left(J+1\right)}/T\right)-1}
   {\exp\left(\mu\sqrt{J\left(J+1\right)}/T\right)+1}
  \right).
\end{eqnarray}
In the analogy with the large $N$ reduced model on a flat
background, we find that
\begin{equation}
 \hat{W}_{(0)}^{\rm 1-loop}
  =\sum_{s}
  \sum_{J,M,\tilde{M}}
  8\log
  \left(
   \frac{\exp\left(\mu\sqrt{J\left(J+1\right)}/T\right)-1}
   {\exp\left(\mu\sqrt{J\left(J+1\right)}/T\right)+1}
  \right),
\end{equation}
where $\tilde{M}={1\over 2}(s-t)$. 
We have introduced a cutoff such that $s<2\Lambda$, so that the maximal value of
$J$ and $\tilde{M}$ are $N_{0}$ and $\Lambda$, respectively.
Then we separate the summation over $J$ into two parts at the value $\Lambda$
as follows:
\begin{eqnarray}
 &&\sum_{s=1}^{2\Lambda}
  \sum_{J=0}^{\Lambda}
  \sum_{M=-J}^{J}
  \sum_{\tilde{M}=-J}^{J}
  8\log
  \left(
   \frac{\exp\left(\mu\sqrt{J(J+1)}/T\right)-1}
   {\exp\left(\mu\sqrt{J(J+1)}/T\right)+1}
  \right)
  \nonumber \\
 &&\hspace{10mm}
  +\sum_{s=1}^{2\Lambda}
  \sum_{J=\Lambda+1/2}^{N_{0}}
  \sum_{M=-J}^{J}
  \sum_{\tilde{M}=-\Lambda}^{\Lambda}
  8\log
  \left(
   \frac{\exp\left(\mu\sqrt{J(J+1)}/T\right)-1}
   {\exp\left(\mu\sqrt{J(J+1)}/T\right)+1}
  \right).
  \nonumber \\
\end{eqnarray}
The second term in the above expression can be safely neglected since
we assume that:
\begin{equation}
 T \ll \Lambda.
\end{equation}
If we further divide this effective
action by the overall factor $\sum_{s}$, it agrees with that
of the supersymmetric Yang-Mills theory on $S^3$.
In this sense, the plane-wave matrix model is a
large $N$ reduced model of the supersymmetric Yang-Mills theory on $S^3$.

In this way we can obtain that
\begin{eqnarray}
 \hat{W}_{(0)}^{\rm 1-loop}
  &=&\sum_{J=0}^{\infty}
  \sum_{M=-J}^{J}
  \sum_{\tilde{M}=-J}^{J}
  8\log
  \left(
   \frac{\exp\left(\mu\sqrt{J(J+1)}/T\right)-1}
   {\exp\left(\mu\sqrt{J(J+1)}/T\right)+1}
  \right)
  \nonumber \\
  &=&\sum_{J=0}^{\infty}
  8\log
  \left(
   \frac{\exp\left(\mu\sqrt{J(J+1)}/T\right)-1}
   {\exp\left(\mu\sqrt{J(J+1)}/T\right)+1}
  \right)
  \left(2J+1\right)^2
  \nonumber \\
 &=&\sum_{k=0}^{\infty}
  8\log
  \left(
   \frac{\exp\left(\sqrt{k(k+2)}/rT\right)-1}
   {\exp\left(\sqrt{k(k+2)}/rT\right)+1}
  \right)
  \left(k+1\right)^2,
\end{eqnarray}
where we set $k=2J$. 
Here we take the high temperature limit such that the temperature is
much larger than the inverse radius of $S^3$:
\begin{equation}
 T \gg \frac1r.
\end{equation}
Thus, this limit represent a flat space limit.
The summation over $k$ can be  well approximated by the
integrals over:
\begin{equation}
 x=\frac{\sqrt{k(k+2)}}{rT}.
\end{equation}

We can obtain the following equation:
\begin{equation}
 \int_{0}^{\infty}\!dx\,
  8\log\left(
       \frac{{\rm e}^{x}-1}{{\rm e}^{x}+1}
      \right)
  \left(
   r^3T^3x^2
   +\frac12rT
  \right).
\end{equation}
This integral is evaluated as:
\begin{equation}
 \hat{W}_{(0)}^{\rm 1-loop}
  =-\frac{\pi^4}{3}r^3T^3
  -\pi^2rT
  +{\cal O}\left(\frac{1}{T}\right).
\end{equation}
Similarly, we evaluate the sub-leading terms of the one-loop effective action:
\begin{eqnarray}
 &&\hat{W}_{(1)}^{\rm 1-loop}
 =\sum_{s,t}
  \left\{
   \sum_{l}
   \frac{9\mu^2}{4}
   {\rm Tr}
   \frac{1}{\left(2\pi lT\right)^2+\mu^2J\left(J+1\right)}
   -\sum_{h}
   \frac{9\mu^2}{4}
   {\rm Tr}
   \frac{1}{\left(2\pi hT\right)^2+\mu^2J\left(J+1\right)}
   \right.
  \nonumber \\
 &&\hspace{30mm}
    -\sum_{l}
   \frac{\mu^4}{2}
   {\rm Tr}
   J\left(J+1\right)
   \left(
    \frac{1}{\left(2\pi lT\right)^2+\mu^2J\left(J+1\right)}
   \right)^2
   \nonumber \\
 &&\hspace{40mm}
  \left.
   +\sum_{h}
   \frac{\mu^4}{2}
   {\rm Tr}
   J\left(J+1\right)
   \left(
    \frac{1}{\left(2\pi hT\right)^2+\mu^2J\left(J+1\right)}
   \right)^2
   \right\}.
\end{eqnarray}
By taking the high temperature limit such that the temperature is
much larger than the inverse radius of $S^3$,
it can be evaluated as follows:
\begin{equation}
 \hat{W}_{(1)}^{\rm 1-loop}
  =\frac{3\pi^2}{2}rT
  +{\cal O}\left(\frac{1}{T}\right).
\end{equation}

In order to examine to what extent a plane wave matrix model can 
explore the planar sector of super Yang-Mills theory on $S^3$,
we further calculate the two-loop effective action of the plane wave
matrix model at finite temperature.
We describe the detailed calculations of the two-loop effective action
in the appendix. The main conclusion is that the equivalence is valid in the
high temperature limit as
the contributions from the non-planar diagrams can be neglected
in comparison to those from the planar diagrams in such a limit.
The two-loop effective action is given by
\begin{equation}
 \hat{W}^{\rm 2-loop}
  =2\pi^4\frac{g_{\rm PW}^2\mu^2}{N_{0}}r^6T^3.
\end{equation}
Here, recalling the following relation between the coupling constants in section 2:
\begin{equation}
 \lim_{N_{0}\to\infty}
  \frac{2g_{\rm PW}^2}{\mu N_{0}}
  =\frac{1}{16\pi^2}g_{\rm SYM}^2,
\end{equation}
we obtain the following equation:
\begin{equation}
 \hat{W}^{\rm 2-loop}
  =\frac{\pi^2}{2}
  g_{\rm SYM}^2r^3T^3.
\end{equation}

We summarize the effective action of the plane wave matrix model
at a finite temperature up to the two-loop level:
\begin{equation}
 \hat{W}/{\rm Vol}\left(S^3\right)
=
-\frac{\pi^2}{6}T^3
+\frac14g_{\rm SYM}^2T^3
+\frac{1}{4r^2}T
+{\cal O}\left(\frac{1}{T}\right),
\end{equation}
where we have divided the effective action by the volume of $S^3$:
\begin{equation}
 {\rm Vol}\left(S^3\right)=2\pi^2r^3.
\end{equation}
This effective action is equal to
$\beta$ times the free energy density of the ${\cal N}=4$ supersymmetric
Yang-Mills theory on $S^3$
\cite{burgess_constable_myers-1999,b_sundborg-2000,aharony_marsano_minwalla_papadodimas_raamsdonk-2004,yamada_yaffe-2006,harmark_orselli-2006}.

\section{Conclusions and discussions}

In this paper, we have investigated the properties of the ${\cal N}=4$
supersymmetric Yang-Mills theory on $S^3$ at finite temperature by using
the plane wave matrix model.

We have formally derived the action of the ${\cal N}=4$ supersymmetric Yang-Mills
theory on ${\mathbb R} \times S^3$ from the action of the plane wave
matrix model by taking the large $N$ limit.
Forthermore, we have calculated the effective action of the plane wave matrix
model around $S^3$ configuration at the two-loop level.
We have found that the effective action of the plane wave matrix
model agrees with the free energy of the ${\cal N}=4$
supersymmetric Yang-Mills theory on $S^3$ at two-loop level in the high temperature limit.
Therefore, we can conclude that the nonperturbative properties of
${\cal N}=4$ supersymmetric
Yang-Mills theory on $S^3$ at finite temperature can be explored by the
plane wave matrix model.
Our results serve as a nontrivial check that the plane wave matrix model
can be regarded as a large $N$ reduced model of the ${\cal N}=4$ supersymmetric
Yang-Mills theory on ${\mathbb R} \times S^3$.
However the nonplanar contributions at the two loop level 
differ from those on $ S^3$. They are rather of $ S^2$ type since the propagators carry vanishing 
$\tilde{M}$ in these contributions.
They can be neglected only in the high temperature limit.
In this sense a construction of a large $N$ reduced model on a curved manifold ($S^3$
in this case) is successful only in a flat manifold limit.

It is interesting to investigate nonperturbative properties of the ${\cal N}=4$
supersymmetric Yang-Mills theory on $S^1 \times S^3$ in connection to AdS/CFT
correspondence.
This correspondence states that the large $N$ ${\cal
N}=4$ supersymmetric Yang-Mills theory on ${\mathbb R} \times S^3$ at
strong coupling region is solved in terms of the type I\hspace{-.1em}IB
supergravity on $AdS_{5} \times S^5$.
We have shown that the ${\cal N}=4$ supersymmetric Yang-Mills theory on
$S^1 \times S^3$ at weak coupling region is consistent with the plane
wave matrix model at  quantum level.
We hope to evaluate the behavior of the ${\cal N}=4$ supersymmetric Yang-Mills theory
on $S^1 \times S^3$ at strong coupling region by using the plane wave
matrix model.

\section*{Acknowledgments}

We would like to thank H. Kaneko for discussions and valuable comments.
We are also grateful to G. Ishiki, S. Shimasaki and A. Tsuchiya for
discussions and valuable comments.
We thank the Yukawa Institute for Theoretical Physics at Kyoto
University, where this work was initiated during the YITP-W-08-04 on
``Development of Quantum Field Theory and String Theory''.
Y.K. acknowledges the warm hospitality at ICTS of TIFR where a part of this work was carried out.

\appendix

\section{Two-loop effective action of plane wave matrix model}

In this appendix, we calculate the two-loop effective action of the
plane wave matrix model on $S^3$ at finite temperature.
The effective action $\hat{W}$ is evaluated as follows:
\begin{eqnarray}
 \hat{W}
  &=&-\log\!\int\!dx_{i}\,dx_{m}\,dA_{0}\,dc\,db\,d\varphi\;
    {\rm e}^{-\tilde{S}_{\rm PW}}
    \nonumber \\
  &=&\hat{W}^{\rm 1-loop}+\hat{W}^{\rm 2-loop},
\end{eqnarray}
where
\begin{equation}
 \hat{W}^{\rm 2-loop}
  =-\log
  \left(
   \frac{\int\!dx_{i}\,dx_{m}\,dA_{0}\,dc\,db\,d\varphi\,
   {\rm e}^{-\tilde{S}_{\rm PW}^{\rm 2-loop}}
   {\rm e}^{-\tilde{S}_{\rm PW}^{\rm 1-loop}}}
   {\int\!dx_{i}\,dx_{m}\,dA_{0}\,dc\,db\,d\varphi\,
   {\rm e}^{-\tilde{S}_{\rm PW}^{\rm 1-loop}}} 
  \right)
  \equiv
  \left<
   {\rm e}^{-\tilde{S}_{\rm PW}^{\rm 2-loop}}
  \right>_{\rm 1PI},
\end{equation}
and
\begin{eqnarray}
 &&\tilde{S}_{\rm PW}^{\rm 2-loop}
  =\frac{1}{g_{\rm PW}^2\mu^2}
  \int_{0}^{\beta}\!dt
  \sum_{s,t}
  {\rm Tr}
  \Biggl\{
   {\rm i}\left(\partial_{0}x_{i}^{(s,t)}\right)
   \left[A_{0},x^{i}\right]^{(t,s)}
   -\mu \left({\cal L}_{i} A_{0}^{(s,t)}\right)
   \left[A_{0},x^{i}\right]^{(t,s)}
  \nonumber \\
 &&\hspace{10mm}
  +\frac{{\rm i}}{2}\mu
  \epsilon^{ijk}x_{i}^{(s,t)}
  \left[x_{j},x_{k}\right]^{(t,s)}
  +\mu \left({\cal L}_{i}x_{j}^{(s,t)}\right)
  \left[x^{i},x^{j}\right]^{(t,s)}
  +{\rm i}\left(\partial_{0}x_{m}^{(s,t)}\right)
  \left[A_{0},x^{m}\right]^{(t,s)}
  \nonumber \\
 &&\hspace{10mm}
  +\mu \left({\cal L}_{i} x_{m}^{(s,t)}\right)
  \left[x^{i},x^{m}\right]^{(t,s)}
  +{\rm i}\left(\partial_{0}b^{(s,t)}\right)
  \left[A_{0},c\right]^{(t,s)}
  +\mu \left({\cal L}_{i}b^{(s,t)}\right)
  \left[x^{i},c\right]^{(t,s)}
  \nonumber \\
 &&\hspace{10mm}
  -\frac12\bar{\varphi}^{(s,t)}\Gamma^{0}
  \left[A_{0},\varphi\right]^{(t,s)}
  -\frac12\bar{\varphi}^{(s,t)}\Gamma^{i}
  \left[x_{i},\varphi\right]^{(t,s)}
  -\frac12\bar{\varphi}^{(s,t)}\Gamma^{m}
  \left[x_{m},\varphi\right]^{(t,s)}
  \nonumber \\
 &&\hspace{10mm}
  -\frac12\left[A_{0},x_{i}\right]^{(s,t)2}
  -\frac14\left[x_{i},x_{j}\right]^{(s,t)2}
  -\frac12\left[A_{0},x_{m}\right]^{(s,t)2}
  \nonumber \\
 &&\hspace{70mm}
  -\frac12\left[x_{i},x_{m}\right]^{(s,t)2}
  -\frac14\left[x_{m},x_{n}\right]^{(s,t)2}
  \Biggr\}.
\end{eqnarray}
We define $\left<\cdots\right>_{\rm 1PI}$ as a summation over only 1PI
(1-Particle-Irreducible) diagrams.
To simplify the following calculations, we combine the action
$\tilde{S}_{\rm PW}^{\rm 2-loop}$ as follows:
\begin{eqnarray}
 &&\tilde{S}_{\rm PW}^{\rm 2-loop}
  =\frac{1}{g_{\rm PW}^2\mu^2}
  \int_{0}^{\beta}\!dt
  \sum_{s,t}
  {\rm Tr}
  \Biggl\{
   -\left({\cal P}_{I} x_{J}^{(s,t)}\right)
   \left[x^{I},x^{J}\right]^{(t,s)}
   +\frac{{\rm i}}{2}\mu\epsilon^{ijk}
   x_{i}^{(s,t)}
   \left[x_{j},x_{k}\right]^{(t,s)}
   \nonumber \\
 &&\hspace{20mm}
  -\left({\cal P}_{I} b^{(s,t)}\right)
   \left[x^{I},c\right]^{(t,s)}
   -\frac12\bar{\varphi}^{(s,t)}\Gamma^{I}
   \left[x_{I},\varphi\right]^{(t,s)}
   -\frac14\left[x_{I},x_{J}\right]^{(s,t)2}
  \Biggr\},
\end{eqnarray}
where the index $I=0,\cdots,9$, and we set that
\begin{equation}
 p_{0}=-{\rm i}\partial_{0},
  \hspace{5mm}
  p_{i}=-\mu L_{i},
  \hspace{5mm}
  p_{m}=0,
  \hspace{5mm}
  x_{0}=A_{0}.
\end{equation}
Now, there are five 1PI diagrams to evaluate which are illustrated in Fig. 1.
The diagrams (a), (b) and (c) represent the contributions from gauge
fields, and (c) involves the Myers type interaction.
The diagrams (d) and (e) represent the contributions from ghost and
fermion fields respectively.
\begin{figure}[htbp]
 \begin{center}
  \includegraphics[scale=1.0]{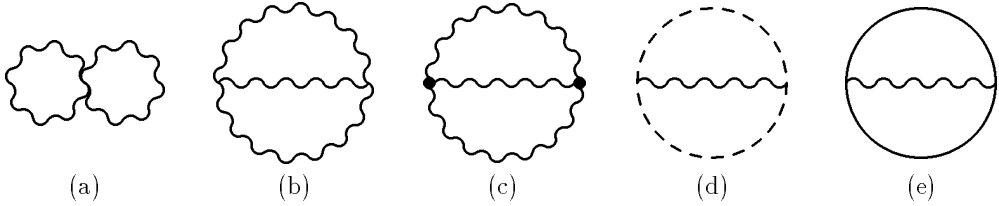}
  \caption{Feynman diagrams of two-loop corrections to the effective
  action \cite{imai_kitazawa_takayama_tomino-2004}.}
  \label{fig:feynman_diagrams}
 \end{center}
\end{figure}
%

\subsection{Propagators}

\subsubsection{Bosonic propagators}

First of all, we derive the bosonic propagators of the plane wave matrix
model.
From the quadratic terms for the gauge fields $x_{i}^{(s,t)}$ in the
action (\ref{Euclidean_S-PW}), we can read out the propagators of gauge boson
modes $x_{i JM}^{(s,t)}$ as follows:
\begin{eqnarray}
 &&\frac{1}{g_{\rm PW}^2\mu^2}
  \int_{0}^{\beta}\!dt\!
  \sum_{s,t}
  {\rm Tr}
  \Biggl\{
   -\frac12 x_{i}^{(s,t)}
   \left(
    -\delta^{ij}\partial_{0}^2
    +\delta^{ij}\mu^2 {\cal L}_{k}^2
    +\mu^2\delta^{ij}
    -{\rm i}\mu^2\epsilon^{ijk}{\cal L}_{k}
   \right)
   x_{j}^{(t,s)}
  \Biggr\}
  \nonumber \\
 &=&\frac{1}{g_{\rm PW}^2\mu^2}
  \int_{0}^{\beta}\!dt\!
  \sum_{s,t}
  {\rm Tr}
  \Biggl\{
   -\frac12
   \sum_{l_{1}}
   \sum_{J_{1},M_{1}}
   {\rm e}^{{\rm i}\omega_{l_{1}}t}
   x_{il_{1}J_{1}M_{1}}^{(s,t)}
   \otimes
   \hat{Y}_{J_{1}M_{1}}^{(j_{s},j_{t})}
   \nonumber \\
 &&\times
  \left(
   \delta^{ij}\omega_{l_{2}}^2
   +\delta^{ij}\mu^2J_{2}\left(J_{2}+1\right)
    +\mu^2\delta^{ij}
    -{\rm i}\mu^2\epsilon^{ijk}{\cal L}_{k}
   \right)
  \sum_{l_{2}}
  \sum_{J_{2},M_{2}}
  {\rm e}^{{\rm i}\omega_{l_{2}}t}
  x_{jl_{2}J_{2}M_{2}}^{(t,s)}
  \otimes
  \hat{Y}_{J_{2}M_{2}}^{(j_{t},j_{s})}
  \Biggr\}
 \nonumber \\
 &&=\frac12
  \sum_{s,t}
  \sum_{l_{1},l_{2}}
  \sum_{J_{1},M_{1}}
  \sum_{J_{2},M_{2}}
  x_{il_{1}J_{1}M_{1}}^{(s,t)}
  (-1)^{M_{1}-\left(j_{s}-j_{t}\right)}
  \frac{-\beta N_{0}}{g_{\rm PW}^2\mu^2}
  \delta_{l_{1}-l_{2}}
  \delta_{J_{1}J_{2}}
  \delta_{M_{1}-M_{2}}
  \nonumber \\
 &&\hspace{20mm}
  \times
  \left(
   \delta^{ij}\omega_{l_{2}}^2
   +\delta^{ij}\mu^2J_{2}\left(J_{2}+1\right)
    +\mu^2\delta^{ij}
    -{\rm i}\mu^2\epsilon^{ijk}{\cal L}_{k}
   \right)
    x_{jl_{2}J_{2}M_{2}}^{(t,s)}.
\end{eqnarray}
Note that the quantum fluctuations are expanded by a plane wave and a fuzzy
spherical harmonics as follows:
\begin{equation}
 x_{i}^{(s,t)}(t)=
  \sum_{l=-\infty}^{\infty}
  \sum_{J=0}^{\infty}
  \sum_{M=-J}^{J}
  {\rm e}^{{\rm i}\omega_{l}t}
  x_{ilJM}^{(s,t)}
  \otimes
  \hat{Y}^{(j_s,j_t)}_{JM}.
\end{equation}
Then, we can obtain the propagator of gauge boson modes as follows: 
\begin{eqnarray}
 &&\Bigl<
  x_{il_{1}J_{1}M_{1}}^{(j_{s},j_{t})}\,
  x_{jl_{2}J_{2}M_{2}}^{(j_{t},j_{s})}
 \Bigr>
 =\frac{g_{\rm PW}^2\mu^2}{\beta N_{0}}
 \frac{(-1)^{M_{1}-\left(j_{s}-j_{t}\right)}}
 {
 \delta^{ij}\omega_{l_{2}}^2
 +\delta^{ij}\mu^2J_{2}\left(J_{2}+1\right)
 +\mu^2\delta^{ij}
 -{\rm i}\mu^2\epsilon^{ijk}{\cal L}_{k}
 }
 \nonumber \\
 &&\hspace{60mm}
  \times
  \delta_{l_{1}-l_{2}}
  \delta_{J_{1}J_{2}}
  \delta_{M_{1}-M_{2}}.
\end{eqnarray}
We expand this propagator into the power series of
$\left(\omega_{l}^2+\mu^2J\left(J+1\right)\right)={\cal P}_{{\rm B}}^2$
as follows:
\begin{eqnarray}
 &&\hspace{-10mm}
  \frac{1}
  {
  \delta^{ij}\omega_{l_{2}}^2
  +\delta^{ij}\mu^2J_{2}\left(J_{2}+1\right)
  +\mu^2\delta^{ij}
  -{\rm i}\mu^2\epsilon^{ijk}{\cal L}_{k}
  }
  \nonumber \\
  &&=\frac{1}{{\cal P}_{{\rm B}}^2}
  \delta^{ij}
  -{\rm i}\mu
  \left(
   \frac{1}{{\cal P}_{{\rm B}}^2}
  \right)^2
  \epsilon^{ijk}{\cal P}_{{\rm B}k}
  -\mu^2
  \left(
   \frac{1}{{\cal P}_{{\rm B}}^2}
  \right)^3
  {\cal P}_{\rm B}^{i}
  {\cal P}_{\rm B}^{j}
  +{\cal O}
  \left(
   \frac{1}{{\cal P}_{{\rm B}}^5}
  \right).
\end{eqnarray}
Therefore, the propagator of gauge boson fields are given by
\begin{eqnarray}
 &&\hspace{-10mm}
  \Bigl<
  x_{i}^{(s,t)}(t_{1})\,
  x_{j}^{(t,s)}(t_{2})
 \Bigr>
 \nonumber \\
 &&=
 \sum_{l_{1},l_{2}}
 \sum_{J_{1},M_{1}}
 \sum_{J_{2},M_{2}}
 {\rm e}^{{\rm i}\omega_{l_{1}}t_{1}}
 {\rm e}^{{\rm i}\omega_{l_{2}}t_{2}}
 \left<
  x_{il_{1}J_{1}M_{1}}^{(s,t)}
  x_{jl_{2}J_{2}M_{2}}^{(t,s)}
 \right>
 \otimes
 \hat{Y}_{J_{1}M_{1}}^{(j_{s},j_{t})}
 \hat{Y}_{J_{2}M_{2}}^{(j_{t},j_{s})}
 \nonumber \\
 &&\sim
  \frac{g_{\rm PW}^2\mu^2}{\beta N_{0}}
  \sum_{l=-\infty}^{\infty}
  \sum_{J=0}^{\infty}
  \sum_{M=-J}^{J}
  \Biggl\{
   \frac{1}{{\cal P}_{{\rm B}}^2}
  \delta_{ij}
  -{\rm i}\mu
  \left(
   \frac{1}{{\cal P}_{{\rm B}}^2}
  \right)^2
  \epsilon_{ijk}{\cal P}_{\rm B}^{k}
  -\mu^2
  \left(
   \frac{1}{{\cal P}_{{\rm B}}^2}
  \right)^3
  {\cal P}_{{\rm B}i}
  {\cal P}_{{\rm B}j}
 \Biggr\}
 \nonumber \\
 &&\hspace{20mm}
  \times
  {\rm e}^{{\rm i}\omega_{l}\left(t_{1}-t_{2}\right)}
  (-1)^{M-\left(j_{s}-j_{t}\right)}
  \hat{Y}_{JM}^{(j_{s},j_{t})}
  \hat{Y}_{J-M}^{(j_{t},j_{s})},
\end{eqnarray}
where we note that $\omega_{-l}=-2\pi lT=-\omega_{l}$.

In the same way, we can read off the other propagators of bosonic fields
as follows:
\begin{eqnarray}
 &&\hspace{-10mm}
  \Bigl<
  x_{m}^{(s,t)}(t_{1})\,
  x_{n}^{(t,s)}(t_{2})
 \Bigr>
 \nonumber \\
 &&\sim
  \frac{g_{\rm PW}^2\mu^2}{\beta N_{0}}
  \sum_{l=-\infty}^{\infty}
  \sum_{J=0}^{\infty}
  \sum_{M=-J}^{J}
  \Biggl\{
   \frac{1}{{\cal P}_{{\rm B}}^2}
   \delta_{mn}
   -\frac{\mu^2}{4}
   \left(
    \frac{1}{{\cal P}_{{\rm B}}^2}
   \right)^2
   \delta_{mn}
  \Biggr\}
  \nonumber \\
 &&\hspace{20mm}
  \times
  {\rm e}^{{\rm i}\omega_{l}\left(t_{1}-t_{2}\right)}
  (-1)^{M-\left(j_{s}-j_{t}\right)}
  \hat{Y}_{JM}^{(j_{s},j_{t})}
  \hat{Y}_{J-M}^{(j_{t},j_{s})},
 \\
 \nonumber \\
 &&\hspace{-10mm}
  \Bigl<
  A_{0}^{(s,t)}(t_{1})\,
  A_{0}^{(t,s)}(t_{2})
 \Bigr>
 \nonumber \\
 &&=\frac{g_{\rm PW}^2\mu^2}{\beta N_{0}}
 \sum_{l=-\infty}^{\infty}
 \sum_{J=0}^{\infty}
 \sum_{M=-J}^{J}
 \frac{1}{{\cal P}_{{\rm B}}^2}
 {\rm e}^{{\rm i}\omega_{l}\left(t_{1}-t_{2}\right)}
 (-1)^{M-\left(j_{s}-j_{t}\right)}
 \hat{Y}_{JM}^{(j_{s},j_{t})}
 \hat{Y}_{J-M}^{(j_{t},j_{s})},
 \\
 \nonumber \\
 &&\hspace{-10mm}
  \Bigl<
  c^{(s,t)}(t_{1})\,
  b^{(t,s)}(t_{2})
 \Bigr>
 \nonumber \\
 &&=\frac{g_{\rm PW}^2\mu^2}{\beta N_{0}}
 \sum_{l=-\infty}^{\infty}
 \sum_{J=0}^{\infty}
 \sum_{M=-J}^{J}
 \frac{1}{{\cal P}_{{\rm B}}^2}
 {\rm e}^{{\rm i}\omega_{l}\left(t_{1}-t_{2}\right)}
 (-1)^{M-\left(j_{s}-j_{t}\right)}
 \hat{Y}_{JM}^{(j_{s},j_{t})}
 \hat{Y}_{J-M}^{(j_{t},j_{s})},
\end{eqnarray}
where we expand the other bosonic fields by the plane-wave and the fuzzy
spherical harmonics as follows:
\begin{eqnarray}
 &&x_{m}^{(s,t)}(t)=
  \sum_{l=-\infty}^{\infty}
  \sum_{J=0}^{\infty}
  \sum_{M=-J}^{J}
  {\rm e}^{{\rm i}\omega_{l}t}
  x_{mlJM}^{(s,t)}
  \otimes
  \hat{Y}^{(j_s,j_t)}_{JM},
  \nonumber \\
 &&A_{0}^{(s,t)}(t)=
  \sum_{l=-\infty}^{\infty}
  \sum_{J=0}^{\infty}
  \sum_{M=-J}^{J}
  {\rm e}^{{\rm i}\omega_{l}t}
  A_{0lJM}^{(s,t)}
  \otimes
  \hat{Y}^{(j_s,j_t)}_{JM},
  \nonumber \\
 &&c^{(s,t)}(t)=
  \sum_{l=-\infty}^{\infty}
  \sum_{J=0}^{\infty}
  \sum_{M=-J}^{J}
  {\rm e}^{{\rm i}\omega_{l}t}
  c^{(s,t)}_{lJM}
  \otimes
  \hat{Y}^{(j_s,j_t)}_{JM},
  \nonumber \\
 &&b^{(s,t)}(t)=
   \sum_{l=-\infty}^{\infty}
   \sum_{J=0}^{\infty}
   \sum_{M=-J}^{J}
   {\rm e}^{{\rm i}\omega_{l}t}
   b^{(s,t)}_{lJM}
   \otimes
   \hat{Y}^{(j_s,j_t)}_{JM}.
\end{eqnarray}
Moreover, we can get the following bosonic propagator from the above
propagators:
\begin{eqnarray}
 &&\hspace{-10mm}
  \left<
  x_{I}^{(s,t)}(t_{1})
  x_{J}^{(t,s)}(t_{2})
 \right>
 \sim
 \frac{g_{\rm PW}^2\mu^2}{\beta N_{0}}
 \sum_{l=-\infty}^{\infty}
 \sum_{J=0}^{\infty}
 \sum_{M=-J}^{J}
 \Biggl\{
 \frac{1}{{\cal P}_{\rm B}^2}
 \delta_{IJ}
 -{\rm i}\mu
 \left(
  \frac{1}{{\cal P}_{\rm B}^2}
 \right)^2
 f_{IJK}{\cal P}_{\rm B}^K
 \nonumber \\
 &&\hspace{20mm}
  +\mu^2\left(\frac{1}{{\cal P}_{\rm B}^2}\right)^3
  \left(
   G_{IJ}{\cal P}_{{\rm B}i}^2
   -H_{IJ}
  \right)
  -\mu^2\left(\frac{1}{{\cal P}_{\rm B}^2}\right)^2G_{IJ}
  -\frac{\mu^2}{4}\left(\frac{1}{{\cal P}_{\rm B}^2}\right)^2I_{IJ}
  \Biggr\}
  \nonumber \\
 &&\hspace{20mm}
  \times
  {\rm e}^{{\rm i}\omega_{l}\left(t_{1}-t_{2}\right)}
  (-1)^{M-\left(j_{s}-j_{t}\right)}
  \hat{Y}_{JM}^{(j_{s},j_{t})}
  \hat{Y}_{JM}^{(j_{t},j_{s})},
\end{eqnarray}
where
\begin{eqnarray}
 &&f_{ijk}=\epsilon_{ijk},
  \hspace{10mm}
  \mbox{other}\;
  f_{IJK}=0,
  \nonumber \\
 &&G_{ij}=\delta_{ij},
  \hspace{10mm}
  \mbox{other}\;
  G_{IJ}=0,
  \nonumber \\
 &&H_{ij}={\cal P}_{{\rm B}i}{\cal P}_{{\rm B}j},
  \hspace{10mm}
  \mbox{other}\;
  H_{IJ}=0,
  \nonumber \\
 &&I_{mn}=\delta_{mn},
  \hspace{10mm}
  \mbox{other}\;
  I_{IJ}=0.
\end{eqnarray}
%

\subsubsection{Fermion propagator}

Next we derive the fermion propagator to read out the quadratic term of
$\varphi$ in the action (\ref{Euclidean_S-PW}) as follows:
\begin{eqnarray}
 &&\hspace{-10mm}
  \frac{1}{g_{\rm PW}^2\mu^2}
  \int_{0}^{\beta}\!dt\!
  \sum_{s,t}
  {\rm Tr}
  \Biggl\{
   -\frac12\bar{\varphi}^{(s,t)}
   \left(
    -{\rm i}\Gamma^{0}\partial_{0}
    -\mu\Gamma^{i}{\cal L}_{i}
    -\frac{3{\rm i}\mu}{4}\Gamma^{123}
   \right)
   \varphi^{(t,s)}
  \Biggr\}
   \nonumber \\
 &&=\frac{1}{g_{\rm PW}^2\mu^2}
  \int_{0}^{\beta}\!dt\!
  \sum_{s,t}
  {\rm Tr}
  \Biggl\{
   -\frac12
   \sum_{h_{1}}
   \sum_{J_{1},M_{1}}
   {\rm e}^{-{\rm i}\omega_{h_{1}}t}
   \bar{\varphi}^{(s,t)}_{h_{1}J_1M_1}
   \otimes
   \hat{Y}_{J_{1}M_{1}}^{(j_s,j_t)\dagger}
   \nonumber \\
 &&\hspace{20mm}
  \times
  \left(
   -{\rm i}\Gamma^{0}\partial_{0}
   -\mu\Gamma^{i}{\cal L}_{i}
   -\frac{3{\rm i}\mu}{4}\Gamma^{123}
  \right)
   \sum_{h_{2}}
   \sum_{J_{2},M_{2}}
   {\rm e}^{{\rm i}\omega_{h_{2}}t}
   \varphi_{h_{2}J_{2}M_{2}}^{(t,s)}
   \otimes
   \hat{Y}_{J_{2}M_{2}}^{(j_t,j_s)}
  \Biggr\}
  \nonumber \\
 &&=\frac12
   \sum_{s,t}
   \sum_{h_{1},h_{2}}
   \sum_{J_{1},M_{1}}
   \sum_{J_{2},M_{2}}
   \bar{\varphi}^{(s,t)}_{h_{1}J_{1}M_{1}}
   \frac{-\beta N_{0}}{g_{\rm PW}^2\mu^2}
   \nonumber \\
 &&\hspace{20mm}
  \times
   \left(\Gamma^{0}\omega_{h_{2}}
    -\mu\Gamma^{i}{\cal L}_{i}
    -\frac{3{\rm i}\mu}{4}\Gamma^{123}
   \right)
  \delta_{h_{1}h_{2}}
  \delta_{J_{1}J_{2}}
  \delta_{M_{1}M_{2}}
  \varphi^{(t,s)}_{h_{2}J_{2}M_{2}},
\end{eqnarray}
where we expanded the quantum fluctuations by a plane-wave and a fuzzy spherical
harmonics as follows:
\begin{equation}
 \varphi^{(s,t)}(t)
  =\sum_{h=-\infty}^{\infty}
  \sum_{J=0}^{\infty}
  \sum_{M=-J}^{J}
  {\rm e}^{{\rm i}\omega_{h}t}
  \varphi^{(s,t)}_{hJM}
  \otimes
  \hat{Y}^{(j_{s},j_{t})}_{JM}.
\end{equation}
Thus, we obtain the propagator of fermion modes as follows:
\begin{eqnarray}
 \Bigl<
  \varphi^{(s,t)}_{h_{1}J_{1}M_{1}}\,
  \bar{\varphi}^{(t,s)}_{h_{2}J_{2}M_{2}}
 \Bigr>
 =\frac{g_{\rm PW}^2\mu^2}{\beta N_0}
 \frac{1}
 {
 \Gamma^{0}\omega_{h_{2}}
 -\mu\Gamma^{i}{\cal L}_{i}
 -\frac{3{\rm i}\mu}{4}\Gamma^{123}
 }
 \delta_{h_{1}h_{2}}
 \delta_{J_{1}J_{2}}
 \delta_{M_{1}M_{2}}.
\end{eqnarray}
Here, we expand this propagator into power series of
$\left(\omega_{h}^2+\mu^2J\left(J+1\right)\right)={\cal P}_{{\rm F}}^2$ as follows:
\begin{eqnarray}
 &&\hspace{-10mm}
  \frac{1}
  {
  \Gamma^{0}\omega_{h_{2}}
  -\mu\Gamma^{i}{\cal L}_{i}
  -\frac{3{\rm i}\mu}{4}\Gamma^{123}
  }
  \nonumber \\
  &&=\frac{1}{{\cal P}_{{\rm F}}^2}
  \Gamma^{I}{\cal P}_{{\rm F}I}
  +\frac{3{\rm i}\mu}{4}
  \frac{1}{{\cal P}_{{\rm F}}^2}
  \Gamma^{123}
  -\frac{{\rm i}}{2}
  \left(
   \frac{1}{{\cal P}_{{\rm F}}^2}
  \right)^2
  \Gamma^{IJ}{\cal F}_{{\rm F}IJ}
  \Gamma^{K}{\cal P}_{{\rm F}K}
  \nonumber \\
 &&\hspace{10mm}
  +\frac{3\mu}{8}
  \left(
   \frac{1}{{\cal P}_{{\rm F}}^2}
  \right)^2
  \Gamma^{IJ}{\cal F}_{{\rm F}IJ}
  \Gamma^{123}
  +\frac{9\mu^2}{16}
  \left(
   \frac{1}{{\cal P}_{{\rm F}}^2}
  \right)^2
  \Gamma^{K}{\cal P}_{{\rm F}K}
  \nonumber \\
 &&\hspace{10mm}
  -\frac14
  \left(
   \frac{1}{{\cal P}_{{\rm F}}^2}
  \right)^3
  \Gamma^{IJ}{\cal F}_{{\rm F}IJ}
  \Gamma^{MN}{\cal F}_{{\rm F}MN}
  \Gamma^{K}{\cal P}_{{\rm F}K}
  +{\cal O}
  \left(
   \frac{1}{{\cal P}_{{\rm F}}^4}
  \right).
  \label{fermion_propagator-expand}
\end{eqnarray}
The third term of (\ref{fermion_propagator-expand}) is that
\begin{eqnarray}
 -\frac{{\rm i}}{2}
  \left(
   \frac{1}{{\cal P}_{{\rm F}}^2}
  \right)^2
  \Gamma^{IJ}{\cal F}_{{\rm F}IJ}
  \Gamma^{K}{\cal P}_{{\rm F}K}
  &=&-\frac{{\rm i}}{2}
  \left(\frac{1}{{\cal P}_{\rm F}^2}\right)^2
  \left(
   -\mu f_{IJK}\Gamma^{IJM}
   {\cal P}_{{\rm F}}^{K}{\cal P}_{{\rm F}M}
   -2{\rm i}\mu^2\Gamma^{I}{\cal P}_{{\rm F}I}
  \right)
  \nonumber \\
 &=&\frac{{\rm i}\mu}{2}
  \left(\frac{1}{{\cal P}_{\rm F}^2}\right)^2
  f_{IJK}\Gamma^{IJM}{\cal P}_{\rm F}^K{\cal P}_{{\rm F}M}
  -\mu^2\left(\frac{1}{{\cal P}_{\rm F}^2}\right)^2
  \Gamma^{I}{\cal P}_{{\rm F}I},
  \nonumber \\
\end{eqnarray}
where we used the multiplication law of the gamma matrices as follows:
\begin{equation}
 \Gamma^{IJ}\Gamma^{K}
  =\Gamma^{IJK}
  -\delta^{IK}\Gamma^{J}
  +\delta^{JK}\Gamma^{I}.
\end{equation}
Moreover, the last term of (\ref{fermion_propagator-expand}) is that
\begin{eqnarray}
 -\frac14
  \left(\frac{1}{{\cal P}_{\rm F}^2}\right)^3
  \Gamma^{IJ}{\cal F}_{{\rm F}IJ}
  \Gamma^{MN}{\cal F}_{{\rm F}MN}
  \Gamma^{K}{\cal P}_{{\rm F}K}
  &=&-\frac14
  \left(\frac{1}{{\cal P}_{\rm F}^2}\right)^3
  \left(
   -4\mu^2\Gamma^{I}{\cal P}_{{\rm F}I}{\cal P}_{\rm F}^2
   +\cdots
  \right)
  \nonumber \\
 &=&\mu^2\left(\frac{1}{{\cal P}_{\rm F}^2}\right)^2
  \Gamma^{I}{\cal P}_{{\rm F}I}
  +\cdots,
\end{eqnarray}
where we also used the multiplication law of the gamma matrices as
follows:
\begin{eqnarray}
 \Gamma^{IJ}\Gamma^{MN}\Gamma^{K}
  &=&\Gamma^{IJMNK}
  -\delta^{IM}\Gamma^{JNK}
  +\delta^{IN}\Gamma^{JMK}
  -\delta^{IK}\Gamma^{JMN}
  +\delta^{JM}\Gamma^{INK}
  \nonumber \\
 &&-\delta^{JN}\Gamma^{IMK}
  +\delta^{JK}\Gamma^{IMN}
  -\delta^{MK}\Gamma^{IJN}
  +\delta^{NK}\Gamma^{IJM}
  \nonumber \\
 &&-\delta^{IM}\delta^{JN}\Gamma^{K}
  +\delta^{IM}\delta^{JK}\Gamma^{N}
  -\delta^{IN}\delta^{JK}\Gamma^{M}
  +\delta^{IN}\delta^{JN}\Gamma^{K}
  \nonumber \\
 &&-\delta^{IK}\delta^{JN}\Gamma^{M}
  +\delta^{IK}\delta^{JM}\Gamma^{N}
  -\delta^{MK}\delta^{JN}\Gamma^{I}
  +\delta^{MK}\delta^{IN}\Gamma^{J}
  \nonumber \\
 &&-\delta^{NK}\delta^{IM}\Gamma^{J}
  +\delta^{NK}\delta^{JM}\Gamma^{I}.
\end{eqnarray}
Thus, we can obtain the following equation:
\begin{eqnarray}
 \frac{1}
  {\Gamma^{I}{\cal P}_{{\rm F}I}
  -\frac{3{\rm i}\mu}{4}\Gamma^{123}
  }
  &=&\frac{1}{{\cal P}_{\rm F}^2}
  \Gamma^{I}{\cal P}_{{\rm F}I}
  +\frac{3{\rm i}\mu}{4}
  \frac{1}{{\cal P}_{\rm F}^2}
  \Gamma^{123}
  +\frac{{\rm i}\mu}{2}
  \left(\frac{1}{{\cal P}_{\rm F}^2}\right)^2
  f_{IJK}\Gamma^{IJM}{\cal P}_{\rm F}^{K}{\cal P}_{{\rm F}M}
  \nonumber \\
 &&-\frac{3\mu^2}{8}
  \left(\frac{1}{{\cal P}_{\rm F}^2}\right)^2
  f_{IJK}\Gamma^{IJ}{\cal P}_{\rm F}^{K}
  \Gamma^{123}
  +\frac{9\mu^2}{16}
  \left(\frac{1}{{\cal P}_{\rm F}^2}\right)^2
  \Gamma^{I}{\cal P}_{{\rm F}I}
  +{\cal O}\left(\frac{1}{{\cal P}_{\rm F}^4}\right).
  \nonumber \\
\end{eqnarray}
Therefore, we obtain the fermion propagator of the plane wave matrix
model as follows:
\begin{eqnarray}
 &&\hspace{-10mm}
  \Bigl<
  \varphi^{(s,t)}(t_{1})\,
  \bar{\varphi}^{(t,s)}(t_{2})
 \Bigr>
 \nonumber \\
 &&=\sum_{h_{1},h_{2}}
 \sum_{J_{1},M_{1}}
 \sum_{J_{2},M_{2}}
 {\rm e}^{{\rm i}\omega_{h_{1}}t_{1}}
 {\rm e}^{-{\rm i}\omega_{h_{2}}t_{2}}
 \left<
  \varphi_{h_{1}J_{1}M_{1}}^{(s,t)}
  \bar{\varphi}_{h_{2}J_{2}M_{2}}^{(t,s)}
 \right>
 \otimes
 \hat{Y}_{J_{1}M_{1}}^{(j_{s},j_{t})}
 \hat{Y}_{J_{2}M_{2}}^{(j_{t},j_{s})\dagger}
 \nonumber \\
 &&\sim
  \frac{g_{\rm PW}^2\mu^2}{\beta N_{0}}
  \sum_{h=-\infty}
  \sum_{J=0}^{\infty}
  \sum_{M=-J}^{J}
  \Biggl\{
   \frac{1}{{\cal P}_{{\rm F}}^2}
  \Gamma^{I}{\cal P}_{{\rm F}I}
  +\frac{3{\rm i}\mu}{4}
  \frac{1}{{\cal P}_{{\rm F}}^2}
  \Gamma^{123}
  +\frac{{\rm i}\mu}{2}
  \left(
   \frac{1}{{\cal P}_{{\rm F}}^2}
  \right)^2
  f_{IJK}
  \Gamma^{IJM}{\cal P}_{{\rm F}}^{K}{\cal P}_{{\rm F}M}
  \nonumber \\
 &&\hspace{30mm}
  -\frac{3\mu^2}{8}
  \left(
   \frac{1}{{\cal P}_{{\rm F}}^2}
  \right)^2
  f_{IJK}
  \Gamma^{IJ}{\cal P}_{{\rm F}}^{K}
  \Gamma^{123}
  +\frac{9\mu^2}{16}
  \left(
   \frac{1}{{\cal P}_{{\rm F}}^2}
  \right)^2
  \Gamma^{I}{\cal P}_{{\rm F}I}
  \Biggr\}
  \nonumber \\
 &&\hspace{65mm}
  \times
  {\rm e}^{{\rm i}\omega_{h}\left(t_{1}-t_{2}\right)}
  (-1)^{M-\left(j_s-j_t\right)}
  \hat{Y}^{(j_s,j_t)}_{JM}
  \hat{Y}^{(j_t,j_s)}_{J-M}.
\end{eqnarray}
%

\subsection{Feynman diagrams}

\subsubsection{Feynman diagram involving four-point gauge boson vertex (a)}

We evaluate the 1PI diagram involving a four-point gauge boson vertex.
The four-point gauge boson vertex is given by
\begin{equation}
 V_{4}
  =\frac{1}{g_{\rm PW}^2\mu^2}
  \int_{0}^{\beta}\!dt
  \sum_{s,t}
  {\rm Tr}
  \Biggl\{
   -\frac14
   \left[x_{I},x_{J}\right]^{(s,t)2}
  \Biggr\}.
\end{equation}
It gives rise to the following contribution:
\begin{eqnarray}
 \left<-V_4\right>_{\rm 1PI}
  &=&\frac{1}{g_{\rm PW}^2\mu^2}
  \int_{0}^{\beta}\!dt\!
  \sum_{s,t}
  {\rm Tr}
  \Biggl\{
   \frac14
   \left[x_{I},x_{J}\right]^{(s,t)2}
  \Biggr\}
   \nonumber \\
 &=&\frac{1}{g_{\rm PW}^2\mu^2}
  \int_{0}^{\beta}\!dt\!
  \sum_{s,t,u,v}
  {\rm Tr}
  \Biggl\{
   \frac12
   \left(
    x_{I}^{(s,t)}x_{J}^{(t,u)}x^{(u,v)I}x^{(v,s)J}
    -x_{I}^{(s,t)}x^{(t,u)I}x_{J}^{(u,v)}x^{(v,s)J}
   \right)
   \Biggr\}.
  \nonumber \\
\end{eqnarray}
We can calculate $\left<-V_{4}\right>_{\rm 1PI}$ by performing the Wick
contraction.
\begin{eqnarray}
 &&\hspace{-5mm}
  \left<-V_{4}\right>_{\rm 1PI}
  \nonumber \\
 &&=\frac{1}{g_{\rm PW}^2\mu^2}
  \int_{0}^{\beta}\!dt\!
  \sum_{s,t,u,v}
  {\rm Tr}
  \Biggl\{
   \frac12
   \left(
    \Bigl<x_{I}^{(s,t)}x_{J}^{(t,u)}\Bigr>
    \Bigl<x^{(u,v)I}x^{(v,s)J}\Bigr>
    +\Bigl<x_{J}^{(s,t)}x^{(t,u)I}\Bigr>
    \Bigl<x^{(u,v)J}x_{I}^{(v,s)}\Bigr>
    \right.
    \nonumber \\
 &&\hspace{20mm}
  \left.
    -\Bigl<x_{I}^{(s,t)}x^{(t,u)I}\Bigr>
    \Bigl<x_{J}^{(u,v)}x^{(v,s)J}\Bigr>
    -\Bigl<x^{(s,t)I}x_{J}^{(t,u)}\Bigr>
    \Bigl<x^{(u,v)J}x_{I}^{(v,s)}\Bigr>
   \right)
   \Biggr\}.
\end{eqnarray}
The leading contribution for the diagram involving four-point gauge
boson vertex is given by
\begin{eqnarray}
 &&\left<-V_{4}\right>_{\rm 1PI}
  =
  \frac{g_{\rm PW}^2\mu^2}{2\beta N_{0}}
  \int_{0}^{\beta}\!dt
  \sum_{s,t,u,v}
  {\rm Tr} \nonumber \\
  &&\times
  \Biggl\{
   \sum_{l_{1}}
   \sum_{J_{1},M_{1}}
   \frac{1}{{\cal P}_{\rm B}^2}
   \delta_{IJ}
   {\rm e}^{{\rm i}\omega_{l_{1}}(t_{1}-t_{2})}
   (-1)^{M_{1}-\left(j_{s}-j_{t}\right)}
   \hat{Y}_{J_{1}M_{1}}^{(j_{s},j_{t})}
   \hat{Y}_{J_{1}-M_{1}}^{(j_{t},j_{u})}
   \nonumber \\
 &&\times
   \sum_{l_{2}}
   \sum_{J_{2},M_{2}}
   \frac{1}{{\cal P}_{\rm B}^2}
   \delta^{IJ}
   {\rm e}^{{\rm i}\omega_{l_{2}}(t_{1}-t_{2})}
   (-1)^{M_{2}-\left(j_{u}-j_{v}\right)}
   \hat{Y}_{J_{2}M_{2}}^{(j_{u},j_{v})}
   \hat{Y}_{J_{2}-M_{2}}^{(j_{v},j_{s})}
   \nonumber \\
 &&+
   \sum_{l_{1}}
   \sum_{J_{1},M_{1}}
   \frac{1}{{\cal P}_{\rm B}^2}
   \delta_{J}^{\,\,\,I}
   {\rm e}^{{\rm i}\omega_{l_{1}}(t_{1}-t_{2})}
   (-1)^{M_{1}-\left(j_{s}-j_{t}\right)}
   \hat{Y}_{J_{1}M_{1}}^{(j_{s},j_{t})}
   \hat{Y}_{J_{1}-M_{1}}^{(j_{t},j_{u})}
   \nonumber \\
 &&\times
   \sum_{l_{2}}
   \sum_{J_{2},M_{2}}
   \frac{1}{{\cal P}_{\rm B}^2}
   \delta_{IJ}
   {\rm e}^{{\rm i}\omega_{l_{2}}(t_{1}-t_{2})}
   (-1)^{M_{2}-\left(j_{u}-j_{v}\right)}
   \hat{Y}_{J_{2}M_{2}}^{(j_{u},j_{v})}
   \hat{Y}_{J_{2}-M_{2}}^{(j_{v},j_{s})}
   \nonumber \\
 &&-   \sum_{l_{1}}
   \sum_{J_{1},M_{1}}
   \frac{1}{{\cal P}_{\rm B}^2}
   \delta_{I}^{\,\,\,I}
   {\rm e}^{{\rm i}\omega_{l_{1}}(t_{1}-t_{2})}
   (-1)^{M_{1}-\left(j_{s}-j_{t}\right)}
   \hat{Y}_{J_{1}M_{1}}^{(j_{s},j_{t})}
   \hat{Y}_{J_{1}-M_{1}}^{(j_{t},j_{u})}
   \nonumber \\
 &&\times
    \sum_{l_{2}}
   \sum_{J_{2},M_{2}}
   \frac{1}{{\cal P}_{\rm B}^2}
   \delta_{J}^{\,\,\,J}
   {\rm e}^{{\rm i}\omega_{l_{2}}(t_{1}-t_{2})}
   (-1)^{M_{2}-\left(j_{u}-j_{v}\right)}
   \hat{Y}_{J_{2}M_{2}}^{(j_{u},j_{v})}
   \hat{Y}_{J_{2}-M_{2}}^{(j_{v},j_{s})}
   \nonumber \\
 &&-   \sum_{l_{1}}
   \sum_{J_{1},M_{1}}
   \frac{1}{{\cal P}_{\rm B}^2}
   \delta^{I}_{\,\,\,J}
   {\rm e}^{{\rm i}\omega_{l_{1}}(t_{1}-t_{2})}
   (-1)^{M_{1}-\left(j_{s}-j_{t}\right)}
   \hat{Y}_{J_{1}M_{1}}^{(j_{s},j_{t})}
   \hat{Y}_{J_{1}-M_{1}}^{(j_{t},j_{u})}
   \nonumber \\
 &&\times
   \sum_{l_{2}}
   \sum_{J_{2},M_{2}}
   \frac{1}{{\cal P}_{\rm B}^2}
   \delta^{J}_{\,\,\,I}
   {\rm e}^{{\rm i}\omega_{l_{2}}(t_{1}-t_{2})}
   (-1)^{M_{2}-\left(j_{u}-j_{v}\right)}
   \hat{Y}_{J_{2}M_{2}}^{(j_{u},j_{v})}
   \hat{Y}_{J_{2}-M_{2}}^{(j_{v},j_{s})}
   \nonumber \\
  &=&\frac12
   \sum_{s,t,u}
   \sum_{l_{1},l_{2}}
   \sum_{J_{1},M_{1}}
   \sum_{J_{2},M_{2}}
   {\rm Tr}
   \sum_{stuv}
   \sum_{pq}
   \sum_{J_1M_1}
   \sum_{J_2M_2}
   \left(\frac{1}{\beta^2N_0}\right)^2
   \frac{(-1)^{M_1+M_2-(j_s-j_t)-(j_u-j_v)}}
        {J_1\left(J_1+1\right)J_2\left(J_2+1\right)}
        \nonumber \\
  &&\times
   \hat{Y}^{(j_sj_t)}_{J_1M_1}\,
   \hat{Y}^{(j_tj_p)}_{J_1-M_1}\,
   \hat{Y}^{(j_uj_v)}_{J_2M_2}\,
   \hat{Y}^{(j_vj_q)}_{J_2-M_2}\,
   \delta_{pu}
   \delta_{qs}.
\end{eqnarray}
Here we have inserted the complete set as follows:
\begin{eqnarray}
 \frac{1}{N_0}{\rm Tr}\sum_{J_3M_3}
  (-1)^{M_3-(j_p-j_q)}\,
  \hat{Y}^{(j_p,j_u)}_{J_3M_3}\,
  \hat{Y}^{(j_q,j_s)}_{J_3-M_3}
  =\delta_{pu}
   \delta_{qs}.
\end{eqnarray}
Therefore, we can get the leading term
\begin{eqnarray}
 \left<-V_{4}\right>_{\rm 1PI}
   \sim-45
   \frac{g_{\rm PW}^2\mu^2}{\beta N_0}
   \sum_{l_{1},l_{2}}
   \sum_{s,t,u}
   \sum_{123}\,
   \hat{\Psi}_{123}^{\dagger}
   \frac{1}{{\cal P}_{\rm B}^2{\cal Q}_{\rm B}^2}
   \hat{\Psi}_{123},
\end{eqnarray}
where ${\cal P}_{\rm B}$, ${\cal Q}_{\rm B}$ and ${\cal R}_{\rm B}$ are defined as
follows:
\begin{eqnarray}
 &&{\cal P}_{I}\hat{Y}_{J_{1}M_{1}}^{(j_{s},j_{t})}
  \equiv
  \left[p_{I},\hat{Y}_{J_{1}M_{1}}^{(j_{s},j_{t})}\right],
  \nonumber \\
 &&{\cal Q}_{I} \hat{Y}_{J_{2}M_{2}}^{(j_{s},j_{t})}
  \equiv
  \left[p_{I},\hat{Y}_{J_{2}M_{2}}^{(j_{s},j_{t})}\right],
  \nonumber \\
 &&{\cal R}_{I} \hat{Y}_{J_{3}M_{3}}^{(j_{s},j_{t})}
  \equiv
  \left[p_{I},\hat{Y}_{J_{3}M_{3}}^{(j_{s},j_{t})}\right].
\end{eqnarray}
We have introduced the following wave function:
\begin{equation}
 \hat{\Psi}_{123}
  \equiv
  \frac{1}{N_{0}}
  {\rm Tr}\,
  \hat{Y}^{(j_s,j_t)}_{J_1M_1}\,
  \hat{Y}^{(j_t,j_u)}_{J_2M_2}\,
  \hat{Y}^{(j_u,j_s)}_{J_3M_3},
\end{equation}
and
\begin{equation}
 \sum_{123}
  \equiv
  \sum_{J_{1}=0}^{\infty}
  \sum_{M_{1}=-J_{1}}^{J_{1}}
  \sum_{J_{2}=0}^{\infty}
  \sum_{M_{2}=-J_{2}}^{J_{2}}
  \sum_{J_{3}=0}^{\infty}
  \sum_{M_{3}=-J_{3}}^{J_{3}}.
\end{equation}
%

\subsubsection{Feynman diagram involving three-point gauge boson vertex (b)}

We evaluate the 1PI diagram involving three-point gauge boson
vertices.
The three-point gauge boson vertex is expressed as follows:
\begin{equation}
 V_{3}
  =\frac{1}{g_{\rm PW}^2\mu^2}
  \int_{0}^{\beta}\!dt
  \sum_{s,t}
  {\rm Tr}
  \Biggl\{
   -\left({\cal P}_{I}x_{J}^{(s,t)}\right)
   \left[x^{I},x^{J}\right]^{(t,s)}
  \Biggr\}.
\end{equation}
We can express the contribution corresponding to the
diagram (b) as follows:
\begin{eqnarray}
 &&\hspace{-10mm}
  \left<\frac12 V_{3}V_{3}\right>_{\rm 1PI}
  \nonumber \\
 &&=\frac12
  \Biggl[
   \frac{1}{g_{\rm PW}^2\mu^2}
   \int_{0}^{\beta}\!dt\!
   \sum_{s,t}
   {\rm Tr}
   \Biggl\{
    \left({\cal P}_{I}x_{J}^{(s,t)}\right)
    \left[x^{I},x^{J}\right]^{(t,s)}
   \Biggr\}
  \Biggr]^2
  \nonumber \\
 &&=\frac12
  \Biggl[
   \frac{1}{g_{\rm PW}^2\mu^2}
   \int_{0}^{\beta}\!dt\!
   \sum_{s,t,u}
   {\rm Tr}
   \Biggl\{
    \left({\cal P}_{I}x_{J}^{(s,t)}\right)
    x^{(t,u)I}
    x^{(u,s)J}
    -\left({\cal P}_{I}x_{J}^{(s,t)}\right)
    x^{(t,u)J}
    x^{(u,s)I}
   \Biggr\}
  \Biggr]^2
  \nonumber \\
 &&=
  \frac{1}{2g_{\rm PW}^4\mu^4}
  \int_{0}^{\beta}\!dt_{1}dt_{2}
  \sum_{s,t,u}
  \Biggl[
  {\rm Tr}
   \Biggl\{
   \left({\cal P}_{I} x_{J}^{(s,t)}\right)
   x^{(t,u)I}x^{(u,s)J}
   \Biggr\}\,
   {\rm Tr}
   \Biggl\{
   \left({\cal P}_{M} x_{N}^{(s,t)}\right)
   x^{(t,u)M}x^{(u,s)N}
   \Biggr\}
   \nonumber \\
 &&\hspace{20mm}
  -{\rm Tr}
  \Biggl\{
  \left({\cal P}_{I}x_{J}^{(s,t)}\right)
  x^{(t,u)I}x^{(u,s)J}
  \Biggr\}\,
  {\rm Tr}
  \Biggl\{
  \left({\cal P}_{M} x_{N}^{(s,t)}\right)
  x^{(t,u)N}x^{(u,s)M}
  \Biggr\}
  \nonumber \\
 &&\hspace{20mm}
  -{\rm Tr}
  \Biggl\{
  \left({\cal P}_{I}x_{J}^{(s,t)}\right)
  x^{(t,u)J}x^{(u,s)I}
  \Biggr\}\,
  {\rm Tr}
  \Biggl\{
  \left({\cal P}_{M}x_{N}^{(s,t)}\right)
  x^{(t,u)M}x^{(u,s)N}
  \Biggr\}
  \nonumber \\
 &&\hspace{20mm}
  \left.
  +{\rm Tr}
  \Biggl\{
  \left({\cal P}_{I} x_{J}^{(s,t)}\right)
  x^{(t,u)J}x^{(u,s)I}
  \Biggr\}\,
  {\rm Tr}
  \Biggl\{
  \left({\cal P}_{M} x_{N}^{(s,t)}\right)
  x^{(t,u)N}x^{(u,s)M}
  \Biggr\}
  \right].
  \label{3point-vertex}
\end{eqnarray}
For example, we calculate the first term of (\ref{3point-vertex}) by
applying Wick's theorem.
\begin{eqnarray}
 &&\hspace{-5mm}
  \frac{1}{2g_{\rm PW}^4\mu^4}
  \int_{0}^{\beta}\!dt_{1}dt_{2}
  \sum_{s,t,u}
  \left[
   {\rm Tr}
   \Biggl\{
   \left({\cal P}_{I} x_{J}^{(s,t)}\right)
   x^{(t,u)I}x^{(u,s)J}
   \Biggr\}\,
   {\rm Tr}
   \Biggl\{
   \left({\cal P}_{M} x_{N}^{(s,t)}\right)
   x^{(t,u)M}x^{(u,s)N}
   \Biggr\}
   \right]
  \nonumber \\
 &&=\frac{1}{2g_{\rm PW}^4\mu^4}
  \int_{0}^{\beta}\!dt_{1}dt_{2}
  \sum_{s,t,u}
  {\rm Tr}\,
  {\rm Tr}
  \nonumber \\
 &&\hspace{3mm}
  \times
  \Biggl\{
  \Bigl<
  \left({\cal P}_{I}x_{J}^{(s,t)}(t_{1})\right)x^{(u,s)N}(t_{2})
  \Bigr>
  \Bigl<x^{(t,u)I}(t_{1})x^{(t,u)M}(t_{2})\Bigr>
  \Bigl<
  x^{(u,s)J}(t_{1})\left({\cal P}_{M}x_{N}^{(s,t)}(t_{2})\right)
  \Bigr>
  \nonumber \\
 &&\hspace{5mm}
  +\Bigl<
  \left({\cal P}_{I}x_{J}^{(s,t)}(t_{1})\right)
  \left({\cal P}_{M}x_{N}^{(u,s)}(t_{2})\right)
  \Bigr>
  \Bigl<x^{(t,u)I}(t_{1})x^{(t,u)N}(t_{2})\Bigr>
  \Bigl<x^{(u,s)J}(t_{1})x^{(s,t)M}(t_{2})\Bigr>
  \nonumber \\
 &&\hspace{5mm}
  +\Bigl<
  \left({\cal P}_{I}x_{J}^{(s,t)}(t_{1})\right)x^{(u,s)M}(t_{2})
  \Bigr>
  \Bigl<
  x^{(t,u)I}(t_{1})\left({\cal P}_{M}x_{N}^{(t,u)}(t_{2})\right)
  \Bigr>
  \Bigl<x^{(u,s)J}(t_{1})x^{(s,t)N}(t_{2})\Bigr>
  \Biggr\}.
  \nonumber \\
\end{eqnarray}
Here, we evaluate the particular contraction which is the first term of
the above equation as follows:
\begin{eqnarray}
 &&\hspace{-5mm}
  \frac{g_{\rm PW}^2\mu^2}{\beta N_{0}}
  \int_{0}^{\beta}\!dt_{1}dt_{2}
  \sum_{s,t,u}
   {\rm Tr}\,
   {\rm Tr}
   \nonumber \\
 &&\hspace{3mm}
  \times
  \Biggl\{
   \sum_{l_{1}}
   \sum_{J_{1},M_{1}}
   \frac{1}{{\cal P}_{\rm B}^2}
   \delta_{J}^{\,\,\,N}
   {\rm e}^{{\rm i}\omega_{l_{1}}(t_{1}-t_{2})}
   (-1)^{M_{1}-\left(j_{s}-j_{t}\right)}
   \left({\cal P}_{{\rm B}I}\hat{Y}_{J_{1}M_{1}}^{(j_{s},j_{t})}\right)
   \hat{Y}_{J_{1}-M_{1}}^{(j_{u},j_{s})}
   \nonumber \\
 &&\hspace{5mm}
  \times
   \sum_{l_{2}}
   \sum_{J_{2},M_{2}}
   \frac{1}{{\cal Q}_{\rm B}^2}
   \delta^{IM}
   {\rm e}^{{\rm i}\omega_{l_{2}}(t_{1}-t_{2})}
   (-1)^{M_{2}-\left(j_{t}-j_{u}\right)}
   \hat{Y}_{J_{2}M_{2}}^{(j_{t},j_{u})}
   \hat{Y}_{J_{2}-M_{2}}^{(j_{t},j_{u})}
   \nonumber \\
 &&\hspace{5mm}
  \times
   \sum_{l_{3}}
   \sum_{J_{3},M_{3}}
   \frac{1}{{\cal R}_{\rm B}^2}
   \delta^{J}_{\,\,\,N}
   {\rm e}^{{\rm i}\omega_{l_{3}}(t_{1}-t_{2})}
   (-1)^{M_{3}-\left(j_{u}-j_{s}\right)}
   \hat{Y}_{J_{3}M_{3}}^{(j_{u},j_{s})}
   \left({\cal R}_{{\rm B}M}\hat{Y}_{J_{3}-M_{3}}^{(j_{s},j_{t})}\right)
   \nonumber \\
 &&\hspace{3mm}
  +   \sum_{l_{1}}
   \sum_{J_{1},M_{1}}
   \frac{1}{{\cal P}_{\rm B}^2}
   \delta_{JN}
   {\rm e}^{{\rm i}\omega_{l_{1}}(t_{1}-t_{2})}
   (-1)^{M_{1}-\left(j_{s}-j_{t}\right)}
   \left({\cal P}_{{\rm B}I}\hat{Y}_{J_{1}M_{1}}^{(j_{s},j_{t})}\right)
   \left({\cal P}_{{\rm B}M}\hat{Y}_{J_{1}-M_{1}}^{(j_{u},j_{s})}\right)
   \nonumber \\
 &&\hspace{5mm}
  \times
    \sum_{l_{2}}
   \sum_{J_{2},M_{2}}
   \frac{1}{{\cal Q}_{\rm B}^2}
   \delta^{IN}
   {\rm e}^{{\rm i}\omega_{l_{2}}(t_{1}-t_{2})}
   (-1)^{M_{2}-\left(j_{t}-j_{u}\right)}
   \hat{Y}_{J_{2}M_{2}}^{(j_{t},j_{u})}
   \hat{Y}_{J_{2}-M_{2}}^{(j_{t},j_{u})}
   \nonumber \\
 &&\hspace{5mm}
  \times
   \sum_{l_{3}}
   \sum_{J_{3},M_{3}}
   \frac{1}{{\cal R}_{\rm B}^2}
   \delta^{JM}
   {\rm e}^{{\rm i}\omega_{l_{3}}(t_{1}-t_{2})}
   (-1)^{M_{3}-\left(j_{u}-j_{s}\right)}
   \hat{Y}_{J_{3}M_{3}}^{(j_{u},j_{s})}
   \hat{Y}_{J_{3}-M_{3}}^{(j_{s},j_{t})}
   \nonumber \\
 &&\hspace{3mm}
  +   \sum_{l_{1}}
   \sum_{J_{1},M_{1}}
   \frac{1}{{\cal P}_{\rm B}^2}
   \delta_{J}^{\,\,\,M}
   {\rm e}^{{\rm i}\omega_{l_{1}}(t_{1}-t_{2})}
   (-1)^{M_{1}-\left(j_{s}-j_{t}\right)}
   \left({\cal P}_{{\rm B}I}\hat{Y}_{J_{1}M_{1}}^{(j_{s},j_{t})}\right)
   \hat{Y}_{J_{1}-M_{1}}^{(j_{u},j_{s})}
   \nonumber \\
 &&\hspace{5mm}
  \times
    \sum_{l_{2}}
   \sum_{J_{2},M_{2}}
   \frac{1}{{\cal Q}_{\rm B}^2}
   \delta^{I}_{\,\,\,N}
   {\rm e}^{{\rm i}\omega_{l_{2}}(t_{1}-t_{2})}
   (-1)^{M_{2}-\left(j_{t}-j_{u}\right)}
   \hat{Y}_{J_{2}M_{2}}^{(j_{t},j_{u})}
   \left({\cal P}_{{\rm B}M}\hat{Y}_{J_{2}-M_{2}}^{(j_{t},j_{u})}\right)
   \nonumber \\
 &&\hspace{5mm}
  \times
   \sum_{l_{3}}
   \sum_{J_{3},M_{3}}
   \frac{1}{{\cal R}_{\rm B}^2}
   \delta^{JN}
   {\rm e}^{{\rm i}\omega_{l_{3}}(t_{1}-t_{2})}
   (-1)^{M_{3}-\left(j_{u}-j_{s}\right)}
   \hat{Y}_{J_{3}M_{3}}^{(j_{u},j_{s})}
   \hat{Y}_{J_{3}-M_{3}}^{(j_{s},j_{t})}
   \Biggr\}
   \nonumber \\
 &&=\frac12
 \frac{g_{\rm PW}^2\mu^2}{\beta N_{0}}
   \sum_{s,t,u}
  \sum_{l_{1},l_{2}}
  \sum_{J_{1},M_{1}}
  \sum_{J_{2},M_{2}}
  \sum_{J_{3},M_{3}}
  \frac{1}{N_{0}}
  {\rm Tr}
  \frac{1}{N_{0}}
  {\rm Tr}
  \nonumber \\
 &&\hspace{8mm}
  \times
  \Biggl\{
   \frac{-10{\cal P}_{\rm B}\cdot{\cal R}_{\rm B}}
   {{\cal P}_{\rm B}^2{\cal Q}_{\rm B}^2{\cal R}_{\rm B}^2}
   (-1)^{M_{1}-\left(j_{s}-j_{t}\right)}
   \hat{Y}_{J_{1}M_{1}}^{(j_{s},j_{t})}
   \hat{Y}_{J_{1}-M_{1}}^{(j_{u},j_{s})}
   \nonumber \\
 &&\hspace{10mm}
  +
   \frac{-{\cal P}_{\rm B}^2}
   {{\cal P}_{\rm B}^2{\cal Q}_{\rm B}^2{\cal R}_{\rm B}^2}
   \delta^{IM}
   (-1)^{M_{2}-\left(j_{t}-j_{u}\right)}
   \hat{Y}_{J_{2}M_{2}}^{(j_{t},j_{u})}
   \hat{Y}_{J_{2}-M_{2}}^{(j_{t},j_{u})}
   \nonumber \\
 &&\hspace{10mm}
  +
   \frac{-{\cal P}_{\rm B}\cdot{\cal Q}_{\rm B}}
   {{\cal P}_{\rm B}^2{\cal Q}_{\rm B}^2{\cal R}_{\rm B}^2}
   (-1)^{M_{3}-\left(j_{u}-j_{s}\right)}
   \hat{Y}_{J_{3}M_{3}}^{(j_{u},j_{s})}
   \hat{Y}_{J_{3}-M_{3}}^{(j_{s},j_{t})}
  \Biggr\},
\end{eqnarray}
where we used the following relation:
\begin{equation}
 \sum_{M=-J}^{J}\,
  (-1)^{M-\left(j_s-j_t\right)}
  \hat{Y}^{(j_s,j_t)}_{JM}
  {\cal P}_{I}
  \hat{Y}^{(j_t,j_s)}_{J-M}
  =-\sum_{M=-J}^{J}\,
  (-1)^{M-\left(j_s-j_t\right)}
  \left(
   {\cal P}_{I}
   \hat{Y}^{(j_s,j_t)}_{JM}
  \right)
  \hat{Y}^{(j_t,j_s)}_{J-M}.
  \label{sign-reversal}
\end{equation}
We can obtain the following compact equation to calculate the other
contractions:
\begin{equation}
 \left<\frac12 V_{3}V_{3}\right>_{\rm 1PI}
  \sim\frac{9}{2}
  \frac{g_{\rm PW}^2\mu^2}{\beta N_{0}}
  \sum_{s,t,u}
  \sum_{l_{1},l_{2}}
  \sum_{123}
  \hat{\Psi}_{123}^{\dagger}\,
  \frac{
  2{\cal P}_{\rm B}^2
  -{\cal P}_{\rm B}\cdot{\cal Q}_{\rm B}
  -{\cal P}_{\rm B}\cdot{\cal R}_{\rm B}
  }
  {{\cal P}_{\rm B}^2
  {\cal Q}_{\rm B}^2
  {\cal R}_{\rm B}^2
  }\,
  \hat{\Psi}_{123}.
\end{equation}
Moreover, we use the following relation:
\begin{equation}
 {\cal P}_{\rm B}\cdot{\cal Q}_{\rm B}
  \hat{\Psi}_{123}
  =\frac12
  \left(
   {\cal R}_{\rm B}^2
   -{\cal P}_{\rm B}^2
   -{\cal Q}_{\rm B}^2
  \right)
  \hat{\Psi}_{123},
  \label{momenta-relation}
\end{equation}
and the momentum conserved relation:
\begin{equation}
 {\cal P}_{\rm B}+{\cal Q}_{\rm B}+{\cal R}_{\rm B}=0.
  \label{momenta-conservation}
\end{equation}
Therefore, we can simplify as follows:
\begin{equation}
 \left<\frac12 V_{3}V_{3}\right>_{\rm 1PI}
  \sim\frac{27}{2}
  \frac{g_{\rm PW}^2\mu^2}{\beta N_{0}}
  \sum_{s,t,u}
  \sum_{l_{1},l_{2}}
  \sum_{123}
  \hat{\Psi}_{123}^{\dagger}
  \frac{1}
  {{\cal P}_{\rm B}^2
  {\cal Q}_{\rm B}^2
  }
  \hat{\Psi}_{123}.
\end{equation}
%

\subsubsection{Feynman diagram involving the ghost interactions (d)}

We evaluate the 1PI diagram involving the ghost interactions.
The ghost vertex is expressed as follows:
\begin{equation}
 V_{\rm gh}
  =\frac{1}{g_{\rm PW}^2\mu^2}
  \int_{0}^{\beta}\!dt
  \sum_{s,t}
  {\rm Tr}
  \Biggl\{
   -\left({\cal P}_{I}b^{(s,t)}\right)
   \left[x^{I},c\right]^{(t,s)}
  \Biggr\}.
\end{equation}
We can express the contribution corresponding to the diagram (d) as
follows:
\begin{eqnarray}
  &&\hspace{-10mm}
   \left<\frac12 V_{\rm gh}V_{\rm gh}\right>_{\rm 1PI}
   \nonumber \\
 &&
  =\frac12
  \Biggl[
   \frac{1}{g_{\rm PW}^2\mu^2}
   \int_{0}^{\beta}\!dt\!
   \sum_{s,t}
   {\rm Tr}
   \Biggl\{
    \left({\cal P}_{I}b^{(s,t)}\right)
    \left[x^{I},c\right]^{(t,s)}
   \Biggr\}
  \Biggr]^2
  \nonumber \\
 &&=\frac12
  \Biggl[
   \frac{1}{g_{\rm PW}^2\mu^2}
   \int_{0}^{\beta}\!dt\!
   \sum_{s,t,u}
   {\rm Tr}
   \Biggl\{
    \left({\cal P}_{I}b^{(s,t)}\right)
    x^{(t,u)I}
    c^{(u,s)}
    -\left({\cal P}_{I}b^{(s,t)}\right)
    c^{(t,u)}
    x^{(u,s)I}
   \Biggr\}
  \Biggr]^2
  \nonumber \\
 &&=\frac{1}{2g_{\rm PW}^4\mu^4}
  \int_{0}^{\beta}\!dt_{1}dt_{2}
  \sum_{s,t,u}
  \Biggl[
   {\rm Tr}
   \Biggl\{
   \left({\cal P}_{I} b^{(s,t)}\right)
   x^{(t,u)I}c^{(u,s)}
   \Biggr\}\,
   {\rm Tr}
   \Biggl\{
   \left({\cal P}_{J} b^{(s,t)}\right)
   x^{(t,u)J}c^{(u,s)}
   \Biggr\}
   \nonumber \\
 &&\hspace{20mm}
  -{\rm Tr}
  \Biggl\{
  \left({\cal P}_{I}b^{(s,t)}\right)
  x^{(t,u)I}c^{(u,s)}
  \Biggr\}\,
  {\rm Tr}
  \Biggl\{
  \left({\cal P}_{J} b^{(s,t)}\right)
  c^{(t,u)}x^{(u,s)J}
  \Biggr\}
  \nonumber \\
 &&\hspace{20mm}
  -{\rm Tr}
  \Biggl\{
  \left({\cal P}_{I}b^{(s,t)}\right)
  c^{(t,u)}x^{(u,s)I}
  \Biggr\}\,
  {\rm Tr}
  \Biggl\{
  \left({\cal P}_{J}b^{(s,t)}\right)
  x^{(t,u)J}c^{(u,s)}
  \Biggr\}
  \nonumber \\
 &&\hspace{20mm}
  +{\rm Tr}
  \Biggl\{
  \left({\cal P}_{I} b^{(s,t)}\right)
  c^{(t,u)}x^{(u,s)I}
  \Biggr\}\,
  {\rm Tr}
  \Biggl\{
  \left({\cal P}_{J} b^{(s,t)}\right)
  c^{(t,u)}x^{(u,s)J}
  \Biggr\}
  \Biggr].
  \label{ghost-int}
\end{eqnarray}
For example, we calculate the first term of (\ref{ghost-int}) by
applying Wick's theorem.
\begin{eqnarray}
 &&\hspace{-5mm}
  \frac{1}{2g_{\rm PW}^4\mu^4}
  \int_{0}^{\beta}\!dt_{1}dt_{2}
  \sum_{s,t,u}
  \left[
   {\rm Tr}
   \Biggl\{
   \left({\cal P}_{I} b^{(s,t)}\right)
   x^{(t,u)I}c^{(u,s)}
   \Biggr\}\,
   {\rm Tr}
   \Biggl\{
   \left({\cal P}_{J} b^{(s,t)}\right)
   x^{(t,u)J}c^{(u,s)}
   \Biggr\}
   \right]
  \nonumber \\
 &&=\frac{1}{2g_{\rm PW}^4\mu^4}
  \int_{0}^{\beta}\!dt_{1}dt_{2}
  \sum_{s,t,u}
  {\rm Tr}\,
  {\rm Tr}
  \nonumber \\
 &&\hspace{3mm}
  \times
  \Biggl\{
  \Bigl<
  \left({\cal P}_{I}b^{(s,t)}(t_{1})\right)c^{(u,s)}(t_{2})
  \Bigr>
  \Bigl<x^{(t,u)I}(t_{1})x^{(t,u)J}(t_{2})\Bigr>
  \Bigl<
  c^{(u,s)}(t_{1})\left({\cal P}_{J}b^{(s,t)}(t_{2})\right)
  \Bigr>
  \Biggr\}.
  \nonumber \\
 &&=\frac{1}{2g_{\rm PW}^4\mu^4}
  \int_{0}^{\beta}\!dt_{1}dt_{2}
  \sum_{s,t,u}
   {\rm Tr}\,
   {\rm Tr}
   \nonumber \\
 &&\hspace{3mm}
  \times
  \Biggl\{
  \frac{g_{\rm PW}^2\mu^2}{\beta N_{0}}
   \sum_{l_{1}}
   \sum_{J_{1},M_{1}}
   \frac{1}{{\cal P}_{\rm B}^2}
   {\rm e}^{{\rm i}\omega_{l_{1}}(t_{1}-t_{2})}
   (-1)^{M_{1}-\left(j_{s}-j_{t}\right)}
   \left({\cal P}_{{\rm B}I}\hat{Y}_{J_{1}M_{1}}^{(j_{s},j_{t})}\right)
   \hat{Y}_{J_{1}-M_{1}}^{(j_{u},j_{s})}
   \nonumber \\
 &&\hspace{5mm}
  \times
  \frac{g_{\rm PW}^2\mu^2}{\beta N_{0}}
   \sum_{l_{2}}
   \sum_{J_{2},M_{2}}
   \frac{1}{{\cal Q}_{\rm B}^2}
   \delta^{IJ}
   {\rm e}^{{\rm i}\omega_{l_{2}}(t_{1}-t_{2})}
   (-1)^{M_{2}-\left(j_{t}-j_{u}\right)}
   \hat{Y}_{J_{2}M_{2}}^{(j_{t},j_{u})}
   \hat{Y}_{J_{2}-M_{2}}^{(j_{t},j_{u})}
   \nonumber \\
 &&\hspace{5mm}
  \times
  \frac{g_{\rm PW}^2\mu^2}{\beta N_{0}}
   \sum_{l_{3}}
   \sum_{J_{3},M_{3}}
   \frac{1}{{\cal R}_{\rm B}^2}
   {\rm e}^{{\rm i}\omega_{l_{3}}(t_{1}-t_{2})}
   (-1)^{M_{3}-\left(j_{u}-j_{s}\right)}
   \hat{Y}_{J_{3}M_{3}}^{(j_{u},j_{s})}
   \left({\cal R}_{{\rm B}M}\hat{Y}_{J_{3}-M_{3}}^{(j_{s},j_{t})}\right)
   \Biggr\}
   \nonumber \\
 &&=-\frac12
  \frac{g_{\rm PW}^2\mu^2}{\beta N_{0}}
  \sum_{s,t,u}
  \sum_{l_{1},l_{2}}
  \sum_{J_{1},M_{1}}
  \sum_{J_{2},M_{2}}
  \sum_{J_{3},M_{3}}
  \frac{1}{N_{0}}
  {\rm Tr}
  \frac{1}{N_{0}}
  {\rm Tr}
  \nonumber \\
 &&\hspace{8mm}
  \times
  \Biggl\{
   \frac{1}{{\cal P}_{\rm B}^2}
   (-1)^{M_{1}-\left(j_{s}-j_{t}\right)}
   \left({\cal P}_{{\rm B}I}\hat{Y}_{J_{1}M_{1}}^{(j_{s},j_{t})}\right)
   \hat{Y}_{J_{1}-M_{1}}^{(j_{u},j_{s})}
   \nonumber \\
 &&\hspace{10mm}
  \times
   \frac{1}{{\cal Q}_{\rm B}^2}
   \delta^{IJ}
   (-1)^{M_{2}-\left(j_{t}-j_{u}\right)}
   \hat{Y}_{J_{2}M_{2}}^{(j_{t},j_{u})}
   \hat{Y}_{J_{2}-M_{2}}^{(j_{t},j_{u})}
   \nonumber \\
 &&\hspace{10mm}
  \times
   \frac{1}{{\cal R}_{\rm B}^2}
   (-1)^{M_{3}-\left(j_{u}-j_{s}\right)}
   \left({\cal R}_{{\rm B}J}\hat{Y}_{J_{3}M_{3}}^{(j_{u},j_{s})}\right)
   \hat{Y}_{J_{3}-M_{3}}^{(j_{s},j_{t})}
  \Biggr\},
\end{eqnarray}
where we also used the relation (\ref{sign-reversal}).
We obtain the following equation to evaluate the other contractions:
\begin{eqnarray}
 \left<\frac12 V_{\rm gh}V_{\rm gh}\right>_{\rm 1PI}
  &\sim&\frac12
  \frac{g_{\rm PW}^2\mu^2}{\beta N_{0}}
  \sum_{s,t,u}
  \sum_{l_{1},l_{2}}
  \sum_{123}\,
  \hat{\Psi}_{123}^{\dagger}\,
  \frac{
  {\cal P}_{\rm B}\cdot{\cal Q}_{\rm B}
  +{\cal P}_{\rm B}\cdot{\cal R}_{\rm B}
  }
  {{\cal P}_{\rm B}^2
  {\cal Q}_{\rm B}^2
  {\cal R}_{\rm B}^2
  }\,
  \hat{\Psi}_{123}
  \nonumber \\
 &=&\frac{g_{\rm PW}^2\mu^2}{\beta N_{0}}
  \sum_{s,t,u}
  \sum_{l_{1},l_{2}}
  \sum_{123}\,
  \hat{\Psi}_{123}^{\dagger}\,
  \frac{
  {\cal P}_{\rm B}\cdot{\cal Q}_{\rm B}
  }
  {{\cal P}_{\rm B}^2
  {\cal Q}_{\rm B}^2
  {\cal R}_{\rm B}^2
  }\,
  \hat{\Psi}_{123}.
\end{eqnarray}
Moreover, we obtain the following simplified equation to use the
relation (\ref{momenta-relation}) and the conservation law of momenta
(\ref{momenta-conservation}):
\begin{equation}
 \left<\frac12 V_{\rm gh}V_{\rm gh}\right>_{\rm 1PI}
  \sim-\frac{1}{2}
  \frac{g_{\rm PW}^2\mu^2}{\beta N_{0}}
  \sum_{s,t,u}
  \sum_{l_{1},l_{2}}
  \sum_{123}\,
  \hat{\Psi}_{123}^{\dagger}\,
  \frac{1}
  {{\cal P}_{\rm B}^2
  {\cal Q}_{\rm B}^2
  }\,
  \hat{\Psi}_{123}.
\end{equation}
%

\subsubsection{Feynman diagram involving the fermion interaction (e)}

Finally, we evaluate the diagram involving fermion interactions.
The fermion vertex is expressed as follows:
\begin{equation}
 V_{\rm F}
  =\frac{1}{g_{\rm PW}^2\mu^2}
  \int_{0}^{\beta}\!dt
  \sum_{s,t}
  {\rm Tr}
  \Biggl\{
   -\frac12\bar{\varphi}^{(s,t)}\Gamma^{I}
   \left[x_{I},\varphi\right]^{(t,s)}
  \Biggr\}.
\end{equation}
The 1PI diagram involving fermion vertices is calculated as follows:
\begin{eqnarray}
 \left<\frac12 V_{\rm F}V_{\rm F}\right>_{\rm 1PI}
  &=&\frac12
  \left[
   \frac{1}{g_{\rm PW}^2\mu^2}
   \int_{0}^{\beta}\!dt\!
   \sum_{s,t}
   {\rm Tr}
   \Biggl\{
    \frac12
    \bar{\varphi}^{(s,t)}
    \Gamma^{I}
    \Bigl[x_{I},\varphi\Bigr]^{(t,s)}
   \Biggr\}
  \right]^2
  \nonumber \\
  &=&\frac12
   \left[
    \frac{1}{g_{\rm PW}^2\mu^2}
    \int_{0}^{\beta}\!dt\!
    \sum_{s,t,u}
    {\rm Tr}
    \Biggl\{
     \bar{\varphi}^{(s,t)}
     \Gamma^{I}
     x_{I}^{(t,u)}
     \varphi^{(u,s)}
    \Biggr\}
   \right]^2.
\end{eqnarray}
We can perform the Wick contractions:
\begin{eqnarray}
 &&\hspace{-10mm}
  \left<\frac12 V_{\rm F}V_{\rm F}\right>_{\rm 1PI}
  \nonumber \\
 &&
  \sim
  \frac12
  \frac{1}{g_{\rm PW}^4\mu^4}
  \int_{0}^{\beta}\!dt_{1}dt_{2}
  \sum_{s,t,u}
  {\rm Tr}{\rm Tr}
  \nonumber \\
  &&\hspace{5mm}
   \times
   \frac{g_{\rm PW}^2\mu^2}{\beta N_{0}}
   \sum_{h_{1}}
   \sum_{J_{1},M_{1}}
   \left(
    -\frac{1}{{\cal P}_{\rm F}^2}
    \Gamma^I
    {\cal P}_{{\rm F}I}
    -\frac{3{\rm i}\mu}{4}
    \frac{1}{{\cal P}_{{\rm F}}^2}
    \Gamma^{123}
    -\frac{{\rm i}\mu}{2}
    \left(\frac{1}{{\cal P}_{{\rm F}}^2}\right)^2
    f_{IJK}
    \Gamma^{IJM}
    {\cal P}_{\rm F}^K
    {\cal P}_{{\rm F}M}
   \right)
   \Gamma^{A}
   \nonumber \\
  &&\hspace{5mm}
   \times
   (-1)^{M_1-\left(j_{s}-j_{t}\right)}
   {\rm e}^{{\rm i}\omega_{h_{1}}(t_{1}-t_{2})}
   \hat{Y}^{(j_s,j_t)}_{J_1M_1}
   \hat{Y}^{(j_{u},j_{s})}_{J_1-M_1}
   \nonumber \\
  &&\hspace{5mm}
   \times
   \frac{g_{\rm PW}^2\mu^2}{\beta N_{0}}
   \sum_{l_{2}}
   \sum_{J_2,M_2}
   \frac{1}{{\cal P}_{\rm B}^2}
   \delta_{AB}
   (-1)^{M_{2}-\left(j_{t}-j_{u}\right)}
   {\rm e}^{{\rm i}\omega_{l_{2}}(t_{1}-t_{2})}
   \hat{Y}^{(j_t,j_u)}_{J_2M_2}
   \hat{Y}^{(j_{u},j_{t})}_{J_2-M_2}
   \nonumber \\
  &&\hspace{5mm}
   \times
   \frac{g_{\rm PW}^2\mu^2}{\beta N_{0}}
   \sum_{h_{3}}
   \sum_{J_{3},M_{3}}
   \left(
    \frac{1}{{\cal P}_{\rm F}^2}
    \Gamma^P
    {\cal P}_{{\rm F}P}
    +\frac{3{\rm i}\mu}{4}
    \frac{1}{{\cal P}_{{\rm F}}^2}
    \Gamma^{123}
    +\frac{{\rm i}\mu}{2}
    \left(\frac{1}{{\cal P}_{{\rm F}}^2}\right)^2
    f_{PQR}
    \Gamma^{PQS}
    {\cal P}_{\rm F}^R
    {\cal P}_{{\rm F}S}
   \right)
   \Gamma^{B}
   \nonumber \\
  &&\hspace{5mm}
   \times
   (-1)^{M_3-\left(j_{u}-j_{s}\right)}
   {\rm e}^{{\rm i}\omega_{h_{3}}(t_{1}-t_{2})}
   \hat{Y}^{(j_u,j_s)}_{J_3M_3}
   \hat{Y}^{(j_{s},j_{t})}_{J_3-M_3}.
   \nonumber \\
\end{eqnarray}
We can evaluate the traces of products of gamma matrices as in
appendix A.7. We obtain the following result
\begin{equation}
 \left<\frac12 V_{\rm F}V_{\rm F}\right>_{\rm 1PI}
  \sim
  \sum_{123}
  \hat{\Psi}_{123}^{\dagger}
  \left(
   -32
   \frac{g_{\rm PW}^2\mu^2}{\beta N_{0}}
   \sum_{h_{1},h_{2}}
   \frac{1}
   {{\cal P}_{\rm F}^2
   {\cal Q}_{\rm F}^2
   }
   +64
   \frac{g_{\rm PW}^2\mu^2}{\beta N_{0}}
   \sum_{l_{1},h_{2}}
   \frac{1}
   {{\cal P}_{\rm B}^2
   {\cal Q}_{\rm F}^2
   }
   \right)
  \hat{\Psi}_{123}.
\end{equation}
%

\subsection{Two-loop effective action}

\subsubsection{Bosonic two-loop effective action}

We calculate the bosonic two-loop effective action of the plane wave
matrix model as follows:
\begin{eqnarray}
 \hat{W}_{\rm B}^{\rm 2-loop}
  &=&\left<
      -V_{4}
      +\frac12 V_{3}V_{3}
      +\frac12 V_{\rm gh}V_{\rm gh}
     \right>_{\rm 1PI}
  \nonumber \\
 &=&-32
  \frac{g_{\rm PW}^2\mu^2}{\beta N_{0}}
  \sum_{l_{1},l_{2}}
  \sum_{s,t,u}
  \sum_{123}
  \hat{\Psi}_{123}^{\dagger}
  \frac{1}
  {{\cal P}_{\rm B}^2{\cal Q}_{\rm B}^2}
  \hat{\Psi}_{123}
  \nonumber \\
 &=&-32
  \frac{g_{\rm PW}^2\mu^2}{\beta N_{0}^3}
  \sum_{l_{1},l_{2}}
  \sum_{s,t,u}
  \sum_{J_{1},M_{1}}
  \sum_{J_{2},M_{2}}
  \sum_{J_{3},M_{3}}
  {\rm Tr}\,
  \hat{Y}_{J_{1}M_{1}}^{(j_{s},j_{t})\dagger}
  \hat{Y}_{J_{2}M_{2}}^{(j_{t},j_{u})\dagger}
  \hat{Y}_{J_{3}M_{3}}^{(j_{u},j_{s})\dagger}
  \nonumber \\
 &&\times
  \frac{1}
  {
  \left(
   \omega_{l_{1}}^2
   +\mu^2J_{1}\left(J_{1}+1\right)
  \right)
  \left(
   \omega_{l_{2}}^2
   +\mu^2J_{2}\left(J_{2}+1\right)
  \right)
  }
  {\rm Tr}\,
  \hat{Y}_{J_{1}M_{1}}^{(j_{s},j_{t})}
  \hat{Y}_{J_{2}M_{2}}^{(j_{t},j_{u})}
  \hat{Y}_{J_{3}M_{3}}^{(j_{u},j_{s})}.
  \nonumber \\
\end{eqnarray}
Note that on the analogy with the large $N$ reduced model on a flat background.
So we obtain that
\begin{eqnarray}
 &&\hat{W}_{\rm B}^{\rm 2-loop}
  =-32
  \frac{g_{\rm PW}^2\mu^2}{\beta N_{0}}
  \sum_{l_{1},l_{2}}
  \sum_{r=1}^{\infty}
  \sum_{J_{1}=0}^{\infty}
  \sum_{M_{1}=-J_{1}}^{J_{1}}
  \sum_{\tilde{M}_{1}=-J_{1}}^{J_{1}}
  \sum_{J_{2}=0}^{\infty}
  \sum_{M_{2}=-J_{2}}^{J_{2}}
  \sum_{\tilde{M}_{2}=-J_{2}}^{J_{2}}
  \sum_{J_{3}=0}^{\infty}
  \sum_{M_{3}=-J_{3}}^{J_{3}}
  \sum_{\tilde{M}_{3}=-J_{3}}^{J_{3}}
  \nonumber \\
 &&\hspace{30mm}
  \times
  \frac{
  \left(2J_{1}+1\right)
  \left(2J_{2}+1\right)
  \left(2J_{3}+1\right)
  }
  {
  \left(
   \omega_{l_{1}}^2
   +\mu^2J_{1}\left(J_{1}+1\right)
  \right)
  \left(
   \omega_{l_{2}}^2
   +\mu^2J_{2}\left(J_{2}+1\right)
  \right)
  }
  \nonumber \\
 &&\hspace{30mm}
  \times
  \left(
   \begin{array}{ccc}
    J_{1} & J_{2} & J_{3} \\
    \\
    M_{1} & M_{2} & M_{3} \\
   \end{array}
  \right)^2
  \left(
   \begin{array}{ccc}
    J_{1} & J_{2} & J_{3} \\
    \\
    \tilde{M}_{1} & \tilde{M}_{2} & \tilde{M}_{3} \\
   \end{array}
  \right)^2,
  \nonumber \\
\end{eqnarray}
where we define that $p/2=\tilde{M}_{1}$, $q/2=\tilde{M}_{2}$ and
$(-p-q)/2=\tilde{M}_{3}$, and use the following relation:
\begin{eqnarray}
 &&\frac{1}{N_0}{\rm Tr}\,
  \hat{Y}_{J_1M_1}^{(j_{s},j_{t})}
  \hat{Y}_{J_2M_2}^{(j_{t},j_{u})}
  \hat{Y}_{J_3M_3}^{(j_{u},j_{s})}
  \mathop{\longrightarrow}_{N_{0} \to \infty}
  (-1)^{2J_2-2J_3-\tilde{M}_1}
   \sqrt{
   \left(2J_1+1\right)
   \left(2J_2+1\right)
   \left(2J_3+1\right)
   }
  \nonumber \\
 &&\hspace{50mm}
  \times\left(
          \begin{array}{ccc}
           J_1 & J_2 & J_3 \\
               &     & \\
           M_1 & M_2 & M_3 \\
          \end{array}
          \right)\!
    \left(
          \begin{array}{ccc}
                  J_1  &     J_2     &    J_3 \\
                       &             & \\
           \tilde{M_1} & \tilde{M_2} & \tilde{M_3} \\
          \end{array}
          \right).
    \label{3-fuzzy_harmonics}
\end{eqnarray}
We have a cutoff such that $r < 2\Lambda$, so the maximal value of $J$
and $\tilde{M}$ are $N_{0}$ and $\Lambda$, respectively.
Then, we separate the sums over $J$ in to two parts at $\Lambda$.
After dividing the overall factor $\sum_{r}$, we can obtain that
\begin{eqnarray}
 &&\hspace{-10mm}
  -32
  \frac{g_{\rm PW}^2\mu^2}{\beta N_{0}}
  \sum_{l_{1},l_{2}}
  \sum_{J_{1}=0}^{\infty}
  \sum_{M_{1}=-J_{1}}^{J_{1}}
  \sum_{\tilde{M}_{1}=-J_{1}}^{J_{1}}
  \sum_{J_{2}=0}^{\infty}
  \sum_{M_{2}=-J_{2}}^{J_{2}}
  \sum_{\tilde{M}_{2}=-J_{2}}^{J_{2}}
  \sum_{J_{3}=0}^{\infty}
  \sum_{M_{3}=-J_{3}}^{J_{3}}
  \sum_{\tilde{M}_{3}=-J_{3}}^{J_{3}}
  \nonumber \\
 &&\hspace{-5mm}
  \times
  \frac{
  \left(2J_{1}+1\right)
  \left(2J_{2}+1\right)
  \left(2J_{3}+1\right)
  }
  {
  \left(
   \omega_{l_{1}}^2
   +\mu^2J_{1}\left(J_{1}+1\right)
  \right)
  \left(
   \omega_{l_{2}}^2
   +\mu^2J_{2}\left(J_{2}+1\right)
  \right)
  }
  \left(
   \begin{array}{ccc}
    J_{1} & J_{2} & J_{3} \\
    \\
    M_{1} & M_{2} & M_{3} \\
   \end{array}
  \right)^2
  \left(
   \begin{array}{ccc}
    J_{1} & J_{2} & J_{3} \\
    \\
    \tilde{M}_{1} & \tilde{M}_{2} & \tilde{M}_{3} \\
   \end{array}
  \right)^2
  \nonumber \\
 &&\hspace{-5mm}
  =-32
  \frac{g_{\rm PW}^2\mu^2}{\beta N_{0}}
  \sum_{l_{1},l_{2}}
  \sum_{J_{1}=0}^{\infty}
  \sum_{J_{2}=0}^{\infty}
  \sum_{J_{3}=0}^{\infty}
  \frac{
  \left(2J_{1}+1\right)
  \left(2J_{2}+1\right)
  \left(2J_{3}+1\right)
  }
  {
  \left(
   \omega_{l_{1}}^2
   +\mu^2J_{1}\left(J_{1}+1\right)
  \right)
  \left(
   \omega_{l_{2}}^2
   +\mu^2J_{2}\left(J_{2}+1\right)
  \right)
  }.
  \nonumber \\
\end{eqnarray}
Note that we consider the following cutoff scale: $T \ll \Lambda$.
We thus obtain that
\begin{equation}
 \hat{W}_{\rm B}^{\rm 2-loop}
  =-32
  \frac{g_{\rm PW}^2\mu^2}{\beta N_{0}}
  \sum_{l_{1},l_{2}}
  \sum_{k_{1}=0}^{\infty}
  \sum_{k_{2}=0}^{\infty}
  \sum_{k_{3}=0}^{\infty}
  \frac{
  \left(k_{1}+1\right)
  \left(k_{2}+1\right)
  \left(k_{3}+1\right)
  }
  {
  \left(
   \omega_{l_{1}}^2
   +\frac{\mu^2}{4}k_{1}\left(k_{1}+2\right)
  \right)
  \left(
   \omega_{l_{2}}^2
   +\frac{\mu^2}{4}k_{2}\left(k_{2}+2\right)
  \right)
  },
\end{equation}
where we set that $k_{1}=2J_{1}$, $k_{2}=2J_{2}$ and $k_{3}=2J_{3}$.
The summations over $k_{1}$, $k_{2}$ and $k_{3}$ can be approximated by the integrals over
\begin{equation}
 x_{1}=
  \frac{\sqrt{k_{1}(k_{1}+2)}}{rT},
  \hspace{5mm}
  x_{2}=
  \frac{\sqrt{k_{2}(k_{2}+2)}}{rT},
  \hspace{5mm}
  x_{3}=
  \frac{\sqrt{k_{3}(k_{3}+2)}}{rT}.
\end{equation}
In a high temperature limit, we obtain the
following equation:
\begin{eqnarray}
 &&-32
  \frac{g_{\rm PW}^2\mu^2}{N_{0}}
  \sum_{l_{1},l_{2}}
  \int_{0}^{\infty}\!dx_{1}dx_{2}dx_{3}
  r^2T^2x_{1}
  r^2T^2x_{2}
  r^2T^2x_{3}
  \frac{1}
  {
  \left(
   \left(2\pi l_{1}T\right)^2
   +x_{1}^2T^2
  \right)
  \left(
   \left(2\pi l_{2}T\right)^2
   +x_{2}^2T^2
  \right)
  }
  \nonumber \\
 &&=-32 r^6T^2
  \frac{g_{\rm PW}^2\mu^2}{N_{0}}
  \sum_{l_{1},l_{2}}
  \int_{0}^{\infty}\!dx_{1}dx_{2}dx_{3}
  \frac{x_{1}x_{2}x_{3}}
  {
  \left(
   \left(2\pi l_{1}\right)^2
   +x_{1}^2
  \right)
  \left(
   \left(2\pi l_{2}\right)^2
   +x_{2}^2
  \right)
  }.
  \nonumber \\
\end{eqnarray}
We want to evaluate the sum of the following form:
\begin{equation}
 \sum_{l_{1}=-\infty}^{\infty}
  \sum_{l_{2}=-\infty}^{\infty}
  \frac{x_{1}}
  {\left(2\pi l_{1}\right)^2+x_{1}^2}
  \frac{x_{2}}
  {\left(2\pi l_{2}\right)^2+x_{2}^2}.
  \label{sum-l_{1},l_{2}}
\end{equation}
Since the function $\frac12\coth\left(\frac{z}{2}\right)$ has poles at
$z=2\pi l {\rm i}$ and is everywhere else bounded and analytic,
we may express the equation (\ref{sum-l_{1},l_{2}}) as a contour integral as
follows:
\begin{eqnarray}
 &&\frac{1}{2\pi{\rm i}}
  \oint\!dz_{1}
  \frac{-x_{1}}{z_{1}^2-x_{1}^2}
  \frac12\coth\left(\frac{z_{1}}{2}\right)
  \frac{1}{2\pi{\rm i}}
  \oint\!dz_{2}
  \frac{-x_{2}}{z_{2}^2-x_{2}^2}
  \frac12\coth\left(\frac{z_{2}}{2}\right)
  \nonumber \\
 &&=\frac12 \coth\left(\frac{x_{1}}{2}\right)
  \cdot
  \frac12 \coth\left(\frac{x_{2}}{2}\right).
\end{eqnarray}
Then, with a suitable rearrangement of the exponentials in the
hyperbolic cotangent, we obtain that
\begin{equation}
 \frac14
  +\frac12
  \frac{1}{{\rm e}^{x_{1}}-1}
  +\frac12
  \frac{1}{{\rm e}^{x_{2}}-1}
  +\frac{1}{{\rm e}^{x_{1}}-1}
  \frac{1}{{\rm e}^{x_{2}}-1}
\end{equation}
Therefore, we can get the bosonic two-loop effective action as follows:
\begin{eqnarray}
 &&\hat{W}_{\rm B}^{\rm 2-loop}
  =-32 r^{6}T^{2}
  \frac{g_{\rm PW}^2\mu^2}{N_{0}}
  \int_{0}^{\infty}\!dx_{1}dx_{2}dx_{3}
  x_{3}
  \nonumber \\
 &&\hspace{30mm}
  \times
  \left(
   \frac14
   +\frac12
   \frac{1}{{\rm e}^{x_{1}}-1}
   +\frac12
   \frac{1}{{\rm e}^{x_{2}}-1}
   +\frac{1}{{\rm e}^{x_{1}}-1}
   \frac{1}{{\rm e}^{x_{2}}-1}
  \right).
\end{eqnarray}
%

\subsubsection{Fermionic two-loop effective action}

We calculate the fermionc two-loop effective action of the plane wave
matrix model as follows:
\begin{eqnarray}
 \hat{W}_{\rm F}^{\rm 2-loop}
  &=&\left<\frac12 V_{\rm F}V_{\rm F}\right>_{\rm 1PI}
  \nonumber \\
  &\sim&
  \sum_{123}
  \hat{\Psi}_{123}^{\dagger}
  \left(
   -32
   \frac{g_{\rm PW}^2\mu^2}{\beta N_{0}}
   \sum_{h_{1},h_{2}}
   \frac{1}
   {{\cal P}_{\rm F}^2
   {\cal Q}_{\rm F}^2
   }
   +64
   \frac{g_{\rm PW}^2\mu^2}{\beta N_{0}}
   \sum_{l_{1},h_{2}}
   \frac{1}
   {{\cal P}_{\rm B}^2
   {\cal Q}_{\rm F}^2
   }
   \right)
  \hat{\Psi}_{123}.
\end{eqnarray}

First, we calculate the first term of the fermionic two-loop effective action
as follows: 
\begin{eqnarray}
 \hat{W}_{\rm F(0)}^{\rm 2-loop}
 &=&-32\frac{g_{\rm PW}^2\mu^2}{\beta N_{0}}
  \sum_{h_{1},h_{2}}
  \sum_{s,t,u}
  \sum_{123}
  \hat{\Psi}_{123}^{\dagger}
  \frac{1}
  {{\cal P}_{\rm F}^2{\cal Q}_{\rm F}^2}
  \hat{\Psi}_{123}
  \nonumber \\
 &=&-32
  \frac{g_{\rm PW}^2\mu^2}{\beta N_{0}^3}
  \sum_{h_{1},h_{2}}
  \sum_{s,t,u}
  \sum_{J_{1},M_{1}}
  \sum_{J_{2},M_{2}}
  \sum_{J_{3},M_{3}}
  {\rm Tr}\,
  \hat{Y}_{J_{1}M_{1}}^{(j_{s},j_{t})\dagger}
  \hat{Y}_{J_{2}M_{2}}^{(j_{t},j_{u})\dagger}
  \hat{Y}_{J_{3}M_{3}}^{(j_{u},j_{s})\dagger}
  \nonumber \\
 &&\times
  \frac{1}
  {
  \left(
   \omega_{h_{1}}^2
   +\mu^2J_{1}\left(J_{1}+1\right)
  \right)
  \left(
   \omega_{h_{2}}^2
   +\mu^2J_{2}\left(J_{2}+1\right)
  \right)
  }
  {\rm Tr}\,
  \hat{Y}_{J_{1}M_{1}}^{(j_{s},j_{t})}
  \hat{Y}_{J_{2}M_{2}}^{(j_{t},j_{u})}
  \hat{Y}_{J_{3}M_{3}}^{(j_{u},j_{s})}
  \nonumber \\
\end{eqnarray}
In the same way as the bosonic two-loop effective
action, we obtain that
\begin{equation}
 -32 r^6T^2
  \frac{g_{\rm PW}^2\mu^2}{\beta N_{0}}
  \sum_{h_{1},h_{2}}
  \int_{0}^{\infty}\!dx_{1}dx_{2}dx_{3}
  \frac{x_{1}x_{2}x_{3}}
  {
  \left(
   \left(2\pi h_{1}\right)^2
   +x_{1}^2
  \right)
  \left(
   \left(2\pi h_{2}\right)^2
   +x_{2}^2
  \right)
  }.
\end{equation}
Similarly, we evaluate the sum of the following form:
\begin{equation}
 \sum_{h_{1}=-\infty}^{\infty}
  \sum_{h_{2}=-\infty}^{\infty}
  \frac{x_{1}}
  {\left(2\pi h_{1}\right)^2+x_{1}^2}
  \frac{x_{2}}
  {\left(2\pi h_{2}\right)^2+x_{2}^2}.
  \label{sum-h_{1},h_{2}}
\end{equation}
Since the function $\frac12\tanh\left(\frac{z}{2}\right)$ has poles at
$z=2\pi h {\rm i}$ and is everywhere else bounded and analytic,
we may express the equation (\ref{sum-h_{1},h_{2}}) as a contour integral as
follows:
\begin{eqnarray}
 &&\frac{1}{2\pi{\rm i}}
  \oint\!dz_{1}
  \frac{-x_{1}}{z_{1}^2-x_{1}^2}
  \frac12\tanh\left(\frac{z_{1}}{2}\right)
  \frac{1}{2\pi{\rm i}}
  \oint\!dz_{2}
  \frac{-x_{2}}{z_{2}^2-x_{2}^2}
  \frac12\tanh\left(\frac{z_{2}}{2}\right)
  \nonumber \\
 &&=\frac12 \tanh\left(\frac{x_{1}}{2}\right)
  \cdot
  \frac12 \tanh\left(\frac{x_{2}}{2}\right).
\end{eqnarray}
Then, with a suitable rearrangement of the exponentials in the
hyperbolic tangent, we obtain that
\begin{equation}
 \frac14
  -\frac12
  \frac{1}{{\rm e}^{x_{1}}+1}
  -\frac12
  \frac{1}{{\rm e}^{x_{2}}+1}
  +\frac{1}{{\rm e}^{x_{1}}+1}
  \frac{1}{{\rm e}^{x_{2}}+1}.
\end{equation} 

Then, we calculate the second term of the fermionic two-loop effective action
as follows:
\begin{eqnarray}
 \hat{W}_{\rm F(1)}^{\rm 2-loop}
  &=&\frac{64g_{\rm PW}^2\mu^2}{\beta N_{0}}
  \sum_{l_{1},h_{2}}
  \int_{0}^{\infty}\!dx_{1}dx_{2}dx_{3}
  r^2T^2x_{1}
  r^2T^2x_{2}
  r^2T^2x_{3}
  \nonumber \\
 &&\times
  \frac{1}
  {
  \left(
   \left(2\pi l_{1}T\right)^2
   +x_{1}^2T^2
  \right)
  \left(
   \left(2\pi h_{2}T\right)^2
   +x_{2}^2T^2
  \right)
  }
  \nonumber \\
 &=&\frac{64g_{\rm PW}^2\mu^2}{N_{0}}
  \sum_{l_{1},h_{2}}
  \int_{0}^{\infty}\!dx_{1}dx_{2}dx_{3}
  r^6T^2
  \frac{x_{1}x_{2}x_{3}}
  {
  \left(
   \left(2\pi l_{1}\right)^2
   +x_{1}^2
  \right)
  \left(
   \left(2\pi h_{2}\right)^2
   +x_{2}^2
  \right)
  }.
  \nonumber \\
\end{eqnarray}
Similarly, we calculate the sum of the following form:
\begin{eqnarray}
 &&\hspace{-15mm}
  \sum_{l_{1},h_{2}}
  \frac{x_{1}}
  {
  \left(2\pi l_{1}\right)^2+x_{1}^2
  }
  \frac{x_{2}}
  {
  \left(2\pi h_{2}\right)^2+x_{2}^2
  }
  \nonumber \\
 &=&\frac{1}{2\pi{\rm i}}
  \oint\!dz_{1}
  \frac{-x_{1}}{z_{1}^2-x_{1}^2}
  \frac12\coth\left(\frac{z_{1}}{2}\right)
  \frac{1}{2\pi{\rm i}}
  \oint\!dz_{2}
  \frac{-x_{2}}{z_{2}^2-x_{2}^2}
  \frac12\tanh\left(\frac{z_{2}}{2}\right)
  \nonumber \\
 &=&\frac12\coth\left(\frac{x_{1}}{2}\right)
  \cdot
  \frac12\tanh\left(\frac{x_{2}}{2}\right)
  \nonumber \\
 &=&\frac14
  \left(1+\frac{2}{{\rm e}^{x_{1}}-1}\right)
  \left(1-\frac{2}{{\rm e}^{x_{2}}+1}\right)
  \nonumber \\
 &=&\frac14
  +\frac12
  \frac{1}{{\rm e}^{x_{1}}-1}
  -\frac12
  \frac{1}{{\rm e}^{x_{2}}+1}
  -\frac{1}{{\rm e}^{x_{1}}-1}
  \frac{1}{{\rm e}^{x_{2}}+1}.
\end{eqnarray}
Therefore, we can get the fermionic two-loop effective action as follows:
\begin{eqnarray}
 &&\hat{W}_{\rm F}^{\rm 2-loop}
  =-\frac{32g_{\rm PW}^2\mu^2}{\beta N_{0}}
  r^{6}T^{2}
  \int_{0}^{\infty}\!dx_{1}dx_{2}dx_{3}
  x_{3}
  \nonumber \\
 &&\hspace{30mm}
  \times
  \left(
   \frac14
   -\frac12
   \frac{1}{{\rm e}^{x_{1}}+1}
   -\frac12
   \frac{1}{{\rm e}^{x_{2}}+1}
   +\frac{1}{{\rm e}^{x_{1}}+1}
   \frac{1}{{\rm e}^{x_{2}}+1}
  \right)
  \nonumber \\
 &&\hspace{20mm}
  +\frac{64g_{\rm PW}^2\mu^2}{\beta N_{0}}
  r^{6}T^{2}
  \int_{0}^{\infty}\!dx_{1}dx_{2}dx_{3}
  x_{3}
  \nonumber \\
 &&\hspace{30mm}
  \times
  \left(
   \frac14
   +\frac12
   \frac{1}{{\rm e}^{x_{1}}-1}
   -\frac12
   \frac{1}{{\rm e}^{x_{2}}+1}
   -\frac{1}{{\rm e}^{x_{1}}-1}
   \frac{1}{{\rm e}^{x_{2}}+1}
  \right).
\end{eqnarray}
%

\subsubsection{All contribution of two-loop effective action}
We summarize the two-loop effective action of the plane wave matrix
model at finite temperature as follows:
\begin{eqnarray}
 \hat{W}^{\rm 2-loop}
 &=&\hat{W}_{\rm B}^{\rm 2-loop}+\hat{W}_{\rm F}^{\rm 2-loop}
  \nonumber \\
 &=&-\frac{32g_{\rm PW}^2\mu^2}{\beta N_{0}}
  r^{6}T^{2}
  \int_{0}^{\infty}\!dx_{1}dx_{2}
  \int_{|x_{1}-x_{2}|}^{x_{1}+x_{2}}\!dx_{3}
  x_{3}
  \nonumber \\
 &&\times
  \left(
   \frac{1}{{\rm e}^{x_{1}}-1}
   \frac{1}{{\rm e}^{x_{2}}-1}
   +\frac{1}{{\rm e}^{x_{1}}+1}
   \frac{1}{{\rm e}^{x_{2}}+1}
   +2\frac{1}{{\rm e}^{x_{1}}-1}
   \frac{1}{{\rm e}^{x_{2}}+1}
  \right)
  \nonumber \\
 &=&-\frac{64g_{\rm PW}^2\mu^2}{\beta N_{0}}
  r^{6}T^{2}
  \int_{0}^{\infty}\!dx_{1}dx_{2}
  \nonumber \\
 &&\times
  \left(
   \frac{x_{1}}{{\rm e}^{x_{1}}-1}
   \frac{x_{2}}{{\rm e}^{x_{2}}-1}
   +\frac{x_{1}}{{\rm e}^{x_{1}}+1}
   \frac{x_{2}}{{\rm e}^{x_{2}}+1}
   +2\frac{x_{1}}{{\rm e}^{x_{1}}-1}
   \frac{x_{2}}{{\rm e}^{x_{2}}+1}
  \right)
  \nonumber \\
 &=&-2\pi^4
  \frac{g_{\rm PW}^2\mu^2}{N_{0}}
  r^{6}T^{3}
\end{eqnarray}
%
%

%
\end{document}